\documentclass[aps,prd,reprint,superscriptaddress,nofootinbib,longbibliography]{revtex4-2}

\usepackage{amsmath,amssymb,bm,mathtools}
\usepackage{graphicx}
\usepackage{microtype}
\usepackage{xcolor}
\usepackage[colorlinks=true,linkcolor=blue!60!black,citecolor=blue!60!black,urlcolor=blue!60!black]{hyperref}
\hypersetup{
  pdftitle={The nucleon is not a hard sphere: screening fifth forces
    with Skyrmions},
  pdfsubject={Screened scalar fifth forces and laboratory bounds}
}

\graphicspath{{figures/}}
\allowdisplaybreaks

\newcommand{\dd}{\mathrm{d}}
\newcommand{\mpl}{M_{\mathrm{Pl}}}
\newcommand{\rbg}{\rho_{\mathrm{bg}}}
\newcommand{\rs}{R_{\mathrm{sph}}}
\newcommand{\rhos}{\rho_{\mathrm{sph}}}
\newcommand{\Q}{\mathcal{Q}}
\newcommand{\Mcal}{\mathcal{M}}
\newcommand{\phibg}{\phi_{\mathrm{bg}}}

\newcommand{\order}{\mathcal{O}}

\newcommand{\TraceAmplitudeMeVfm}{112.152}

\newcommand{\ChameleonBetaHalf}{3.35\times 10^{9}}
\newcommand{\ChameleonSphereHalfShift}{14.5}

\newcommand{\PionXNeutron}{0.429}
\newcommand{\PionXProton}{0.784}
\newcommand{\PionCoreNeutronFm}{0.3034}
\newcommand{\PionCoreProtonFm}{0.5541}
\newcommand{\PionChamRatioChiral}{9.7}
\newcommand{\PionChamRatioPhys}{6.9}
\newcommand{\PionSymRatioChiralLow}{650}
\newcommand{\PionSymRatioPhysLow}{26}
\newcommand{\PionSymRatioChiralMid}{65}
\newcommand{\PionSymRatioPhysMid}{14}
\newcommand{\PionSymRatioChiralHigh}{6.5}
\newcommand{\PionSymRatioPhysHigh}{5}

\newcommand{\BenchmarkCoreRadiusFm}{1}
\newcommand{\BenchmarkCoreMassMeV}{939.565}
\newcommand{\BenchmarkBackgroundDensity}{2.32\times10^{-7}}

\newcommand{\NeutronSphereRadiusFm}{0.5}
\newcommand{\ProtonSphereRadiusFm}{0.877}
\newcommand{\ChiralCoreNeutronFm}{0.288675}

\newcommand{\NeutronChameleonBetaRescaling}{0.436790}

\newcommand{\NeutronChameleonBetaHalfEstimate}{1.46\times10^{9}}

\newcommand{\ConstraintQbounceChiralGain}{19}
\newcommand{\ConstraintQbouncePionGain}{16}
\newcommand{\ConstraintQbounceMassMin}{6\times10^{-24}}
\newcommand{\ConstraintQbounceMassMax}{2\times10^{-9}}
\newcommand{\ConstraintQbounceLambdaMax}{1.4\times10^{3}}
\newcommand{\ConstraintInterferometryMassMax}{3\times10^{-5}}
\newcommand{\ConstraintInterferometryLambdaMin}{1.2\times10^{-3}}
\newcommand{\ConstraintHydrogenMassMax}{6.1\times10^{-13}}
\newcommand{\ConstraintHydrogenSphereLambdaMin}{4\times10^{-4}}
\newcommand{\ConstraintHydrogenPionLambdaMin}{1\times10^{-4}}
\newcommand{\ConstraintHydrogenChiralLambdaMin}{1\times10^{-5}}
\newcommand{\ConstraintMuoniumMassMax}{2\times10^{-16}}
\newcommand{\ConstraintInterferometryLambdaMax}{280}
\newcommand{\ConstraintMuoniumWeakEdgeLambdaMin}{1\times10^{-6}}
\newcommand{\ConstraintInterferometryChiralGainSixteen}{0.4}
\newcommand{\ConstraintInterferometryPionGainSixteen}{0.3}
\newcommand{\ConstraintInterferometryChiralGainTwentyFour}{0.8}
\newcommand{\ConstraintInterferometryPionGainTwentyFour}{0.3}
\newcommand{\ConstraintSymInterferometryChiralEv}{1.5}
\newcommand{\ConstraintSymInterferometryPionEv}{1.0}
\newcommand{\ConstraintSymInterferometryChiralKev}{3.0}
\newcommand{\ConstraintSymInterferometryPionKev}{1.4}
\newcommand{\ConstraintSymHydrogenChiralKev}{4.0}
\newcommand{\ConstraintSymHydrogenPionKev}{1.4}
\newcommand{\ConstraintHydrogenDEGainChiral}{2.4}
\newcommand{\ConstraintHydrogenDEGainPion}{1.8}
\newcommand{\ConstraintHydrogenDEMassSphere}{1.7\times10^{-17}}
\newcommand{\ConstraintHydrogenDEMassChiral}{4.5\times10^{-15}}
\newcommand{\ConstraintHydrogenDEMassPion}{1.1\times10^{-15}}
\newcommand{\ConstraintHydrogenDEMassMuonium}{2.1\times10^{-16}}
\newcommand{\ConstraintSymQbounceMassKev}{1.05}
\newcommand{\ConstraintSymQbounceSphereLambda}{2.99\times10^{1}}
\newcommand{\ConstraintSymQbounceChiralLambda}{5.99\times10^{3}}
\newcommand{\ConstraintSymQbouncePionLambda}{5.94\times10^{2}}
\newcommand{\ConstraintSymQbounceChiralGain}{200}
\newcommand{\ConstraintSymQbouncePionGain}{19.8}

\newcommand{\PionValidationCount}{7}

\newcommand{\PionValidationChamMin}{1.14\times10^{-2}}
\newcommand{\PionValidationChamMax}{5.72\times10^{-1}}
\newcommand{\PionValidationSymMassMin}{1.8\times10^{5}}
\newcommand{\PionValidationSymMassMax}{1.35\times10^{8}}

\newcommand{\ConstraintFluxTracePercent}{2.0}
\newcommand{\ConstraintFluxSphereBound}{7\times10^{-6}}

\begin{document}

\title{The nucleon is not a hard sphere: screening fifth forces with
Skyrmions}

\author{Philippe Brax}
\email{philippe.brax@ipht.fr}
\affiliation{Institut de Physique Th\'eorique, Universit\'e Paris-Saclay,
CEA, CNRS, F-91191 Gif-sur-Yvette Cedex, France}

\author{Clare Burrage}
\email{clare.burrage@nottingham.ac.uk}
\affiliation{School of Physics and Astronomy, University of Nottingham,
University Park, Nottingham NG7 2RD, United Kingdom}

\author{Adam Moss}
\email{adam.moss@nottingham.ac.uk}
\affiliation{School of Physics and Astronomy, University of Nottingham,
University Park, Nottingham NG7 2RD, United Kingdom}

\date{\today}

\begin{abstract}
Laboratory searches for chameleon and symmetron fifth forces increasingly
rely on individual nucleons as sources or probes, but model them as
uniform spheres.
Motivated by the Skyrme picture, we instead consider a nucleon profile
with a constant-density core and an $r^{-6}$ pion tail, together with
its exponential truncation at the physical pion mass. We derive
analytic screening charges for the uniform sphere and the $r^{-6}$
profile, and extend the comparison to the exponential truncation
at the physical pion mass using numerical
solutions. Because the outer tail is too dilute to be screened
at any coupling strength, the $r^{-6}$ tail produces a power-law
enhancement over the sphere charge in strong screening, and the
exponential tail has a weaker, logarithmic enhancement. Propagating this difference through the
published analyses, nucleon-sourced exclusion regions extend
systematically further into the strongly coupled regime. For example, symmetron constraints from neutron interferometry at $\mu=1\,\mathrm{keV}$ reach
maximal self-couplings about $\ConstraintSymInterferometryPionKev$ orders of magnitude higher than for a
uniform-sphere neutron ($\ConstraintSymInterferometryChiralKev$ orders of magnitude in the chiral limit),
and at the dark-energy scale the hydrogen $1s$--$2s$ chameleon bound
reaches couplings $\ConstraintHydrogenDEGainPion$ orders of magnitude
weaker than for a uniform-sphere proton ($\ConstraintHydrogenDEGainChiral$
in the chiral limit), overtaking the muonium bound.
\end{abstract}

\maketitle

\section{Introduction}
\label{sec:intro}

Light scalar fields coupled to matter arise in many extensions of the
standard cosmological model \cite{Joyce:2014kja,Koyama:2015vza,Clifton:2011jh,Arza:2026rsl}, including as explanations for dark energy and dark matter. Screening mechanisms, such as the chameleon's
density-dependent mass~\cite{Khoury:2003aq,Khoury:2003rn,Brax:2004qh} and the symmetron's
density-dependent coupling~\cite{Hinterbichler:2010es,Hinterbichler:2011ca} (for earlier related work see Refs.~\cite{Dehnen:1992rr,Gessner:1992flm}), allow such fields
to evade solar-system tests and some laboratory searches.  These models of screening rely on  self-interactions and non-linearities in the scalar field theory, so that the properties of the force mediator depend on the environment. As a result scalar mediated fifth forces  can be dynamically small in experimental searches without the need to fine tune parameters. However, carefully designed experiments can still provide constraints \cite{Burrage:2017qrf,Brax:2021wcv,Mota:2006fz}.

A distinctive class of laboratory probes uses individual
nucleons: gravitational resonance spectroscopy with ultracold neutrons
(qBOUNCE)~\cite{Cronenberg:2018qxf,Fischer:2024eic}, neutron
interferometry~\cite{Fischer:2023eww}, and precision spectroscopy of
hydrogen~\cite{Brax:2022olf}, and these have proved particularly constraining for chameleon and symmetron models with the strongest couplings to matter. In these analyses the nucleon's response is characterized by a
  \emph{screening charge} $\Q\in[0,1]$, which measures the suppression
  of its scalar coupling relative to an unscreened nucleon. The published analyses compute $\Q$ for a uniform sphere of nuclear density representing the matter distribution of neutrons. In this paper, we revisit this hypothesis.

Indeed a nucleon, contrary to the usual assumption made in analysing laboratory experiments involving neutrons, is not a hard sphere.  In this work we use the Skyrme model of baryons~\cite{Skyrme:1961vq,Adkins:1983ya}  to describe a more realistic matter profile.   In the Skyrme
model the trace of the stress-energy
tensor, the source of a conformally coupled scalar, has a localized core
and a power-law tail \cite{Cebulla:2007ei}. Skyrmions are solitons which arise in the description of the Goldstone bosons associated with the breaking of chiral symmetry if four-derivative terms are included in the chiral Lagrangian. Reasonable agreement is found with experimental observations, see for example Refs.~\cite{Cherman:2009gb,Manton:2008rt,Karliner:2015qoa}.  The stability of the solitons, and their finite size,  is a consequence of balancing the two- and four-derivative terms. As a result of the necessarily large four-derivative terms, questions have been asked about the validity of the Skyrme model within effective field theory (EFT), and whether other, higher-order terms should also be included. However,  for the $B=1$ solitons that we consider here the profile is smooth at the center of the Skyrmion and it has been shown that variations in EFT coefficients result in only small changes to the Skyrmion profile \cite{Manton:2024dgh}.

While lattice effective field theories \cite{Lee:2025req} may now provide more detailed insight into the structure of nucleons than the Skyrme model, they do not currently provide a simple profile for the stress-energy of the nucleon state. Therefore we choose to use the Skyrmion profile, as an improvement over the hard sphere profile, to investigate how sensitive constraints on screened models are to nuclear matter distributions.

In what follows we introduce our theoretical model in Sec.~\ref{sec:source}, outlining both the chameleon and symmetron scalar field theories, which we choose as example screening models, and determining the profile of the Skyrmion source. In Secs.~\ref{sec:chameleon} and \ref{sec:symmetron} we derive the screening charges for solid spheres, and Skyrmions of equivalent mass, for  the chameleon and symmetron models respectively. In Sec.~\ref{sec:pion} we determine the changes to these charges when the Skyrmion profile is exponentially truncated at a scale corresponding to the pion mass. In Sec.~\ref{sec:constraints} we show how the constraints of the laboratory experiments, introduced above, change with different descriptions of the neutron. We conclude in Sec.~\ref{sec:discussion}. Appendices provide further details of our derivations. 

In this work we use the $(-,+,+,+)$ metric convention, and set $\hbar = c=1$. $\mpl^2= 1/8\pi G_N$ is the reduced Planck mass squared.

\section{Theoretical setup}
\label{sec:source}

\subsection{Screened scalar fields}
\label{sec:setup-scalars}

Both the chameleon and the symmetron belong to the class of conformally coupled scalars, with
scalar sector
\begin{equation}
 S_\phi=\int\dd^4x\,\sqrt{-g}
 \left[-\frac12(\partial\phi)^2-V(\phi)\right],
 \label{eq:action}
\end{equation}
and matter action
  $S_{\rm m}=S_{\rm m}[\psi_i;A^2(\phi)g_{\mu\nu}]$, where $\psi_i$ represent the matter fields which move on geodesics of the rescaled metric $A^2(\phi)g_{\mu\nu}$.
  When considering continuous matter distributions, the scalar couples to the Einstein-frame matter trace
  $T=g_{\mu\nu}T^{\mu\nu}$ according to
  $\Box\phi=V_{,\phi}-(A_{,\phi}/A)T$.
  For static pressureless matter, writing $\rho=-T/A$ for the
  conserved rest-mass density, the scalar obeys
\begin{equation}
 \nabla^2\phi=V_{,\phi}(\phi)+A_{,\phi}(\phi)\,\rho
 \equiv V_{{\rm eff},\phi}(\phi;\rho),
 \label{eq:eom}
\end{equation}
so the dynamics are governed by the density-dependent effective potential
$V_{\rm eff}=V(\phi)+\rho\,A(\phi)$, and a test particle feels the fifth
force per unit mass
\begin{equation}
 \bm F=-\bm\nabla\ln A(\phi).
 \label{eq:force}
\end{equation}

The chameleon~\cite{Khoury:2003aq} combines an inverse-power potential
with an approximately linear coupling,
\begin{equation}
 V(\phi)=\frac{\Lambda^{n+4}}{\phi^{n}},
 \qquad
 A(\phi)=e^{\phi/M_c}\simeq1+\frac{\phi}{M_c},
 \label{eq:chameleon-model}
\end{equation}
where the energy scale controlling the coupling to matter, $M_c$, can also be expressed in terms of the dimensionless coupling $\beta$ through $M_c=\mpl/\beta$.  The effective potential has a minimum at
$\phi_{\min}(\rho)=(nM_c\Lambda^{n+4}/\rho)^{1/(n+1)}$, where the mass
$m^2=V_{{\rm eff},\phi\phi}$ grows with density. If the corresponding Compton wavelength of the scalar becomes smaller than the size of the object, then such  dense objects develop `thin
shells', where the mass in the interior of the shell  decouples from the exterior field.  The fifth force
is $\bm F\simeq-\bm\nabla\phi/M_c$.

The symmetron~\cite{Hinterbichler:2010es} combines a
$\mathbb{Z}_2$-symmetric Higgs-like potential with a quadratic coupling to matter,
\begin{equation}
 V(\phi)=-\frac{\mu^2}{2}\phi^2+\frac{\lambda}{4}\phi^4,
 \qquad
 A(\phi)=1+\frac{\phi^2}{2M_s^2},
 \label{eq:symmetron-model}
\end{equation}
so the effective quadratic term, in the presence of a non-relativistic perfect fluid, is $(\rho/M_s^2-\mu^2)\,\phi^2/2$.  For
$\rho>\mu^2M_s^2$ the symmetry is restored, $\phi\to0$, and the fifth
force $\bm F\simeq-(\phi/M_s^2)\bm\nabla\phi$ switches off. In vacuum the
symmetry breaks to $\phi_{\rm v}=\mu/\sqrt\lambda$ with mass
$m_{\rm v}=\sqrt2\,\mu$.

In both theories the field has a finite range $m_{\rm bg}^{-1}$, set by
the ambient background density, far larger than a nucleon for the laboratory parameters considered here. Outside a compact source of size $R$, the region $R \ll r \ll m_{\rm bg}^{-1}$, where scalar-potential effects are negligible, is its \emph{near zone}. Here the exterior field is approximately harmonic, with $\phi \simeq \phibg - D/r$. We define the screening charge $\Q\in[0,1]$ as the exterior $1/r$ coefficient
  normalized to its unscreened point-source value. The explicit forms
  for each theory are given in
  Secs.~\ref{sec:chameleon} and~\ref{sec:symmetron}.

\subsection{The $B=1$ Skyrmion source}
\label{sec:skyrmion}

In the Skyrme model~\cite{Skyrme:1961vq,Adkins:1983ya} the nucleon is the
baryon-number-one soliton of the chiral field $U(\bm x)\in SU(2)$, described by the Lagrangian:
\begin{equation}
 \begin{split}
 \mathcal L_{\rm Sk}={}&-\frac{F_\pi^2}{16}\,
 \mathrm{Tr}\,\partial_\mu U\,\partial^\mu U^\dagger\\
 &+\frac{1}{32e^2}\,\mathrm{Tr}[L_\mu,L_\nu][L^\mu,L^\nu]\\
 &+\frac{F_\pi^2m_\pi^2}{16}\,\mathrm{Tr}(U+U^\dagger-2\mathbf1),
 \end{split}
 \label{eq:skyrme}
\end{equation}
where $L_\mu=U^\dagger\partial_\mu U$.
The soliton is described by the hedgehog ansatz
$U=\exp[i\bm\tau\cdot\hat{\bm r}\,F(r)]$ with $F(0)=\pi$, $F(\infty)=0$ \cite{Canfora:2013osa,Manton:2024dgh}.

The nucleon is a field configuration with internal stresses, so the
trace of its stress-energy tensor, which sources the scalar, is not
simply its energy density. Since the Einstein-frame trace is
$T=A^4\tilde T$, with $\tilde T$ the trace computed on the Jordan
metric $A^2g_{\mu\nu}$, the matter term in Eq.~\eqref{eq:eom} is
$-A_{,\phi}A^3\tilde T$. The product $A_{,\phi}\tilde T$ is already the
leading coupling of the scalar to matter, and the dependence of the
Skyrmion profile on $A$ contributes only a relative correction of order
$A-1$. We therefore evaluate the profile at
$A=1$, where the two frames coincide, and define the positive source for the scalar field as
$\rho_{\rm tr}\equiv-\tilde T=\varepsilon-3p$. Here $p$ is the mean of the
three principal pressures. This trace source replaces $\rho$
in Eq.~\eqref{eq:eom}.

For the static hedgehog this source has a nearly constant core of
radius set by $(eF_\pi)^{-1}$, while at large radius the chiral-limit ($m_\pi\to0$) profile falls as $F\to C/r^2$ and the trace, given by the
two-derivative sigma term, decays as $\propto F_\pi^2C^2/r^6$.  We
therefore adopt the compact phenomenological representation
\begin{equation}
 \rho_{\rm tr}(r)=\rho_A
 \begin{cases}
  1, & r<a,\\
  (a/r)^6, & r>a,
 \end{cases}
 \label{eq:trace-source}
\end{equation}
with the amplitude, $\rho_A$,  fixed by the total nucleon mass.

\begin{figure}[tbp]
 \centering
 \includegraphics[width=\columnwidth]{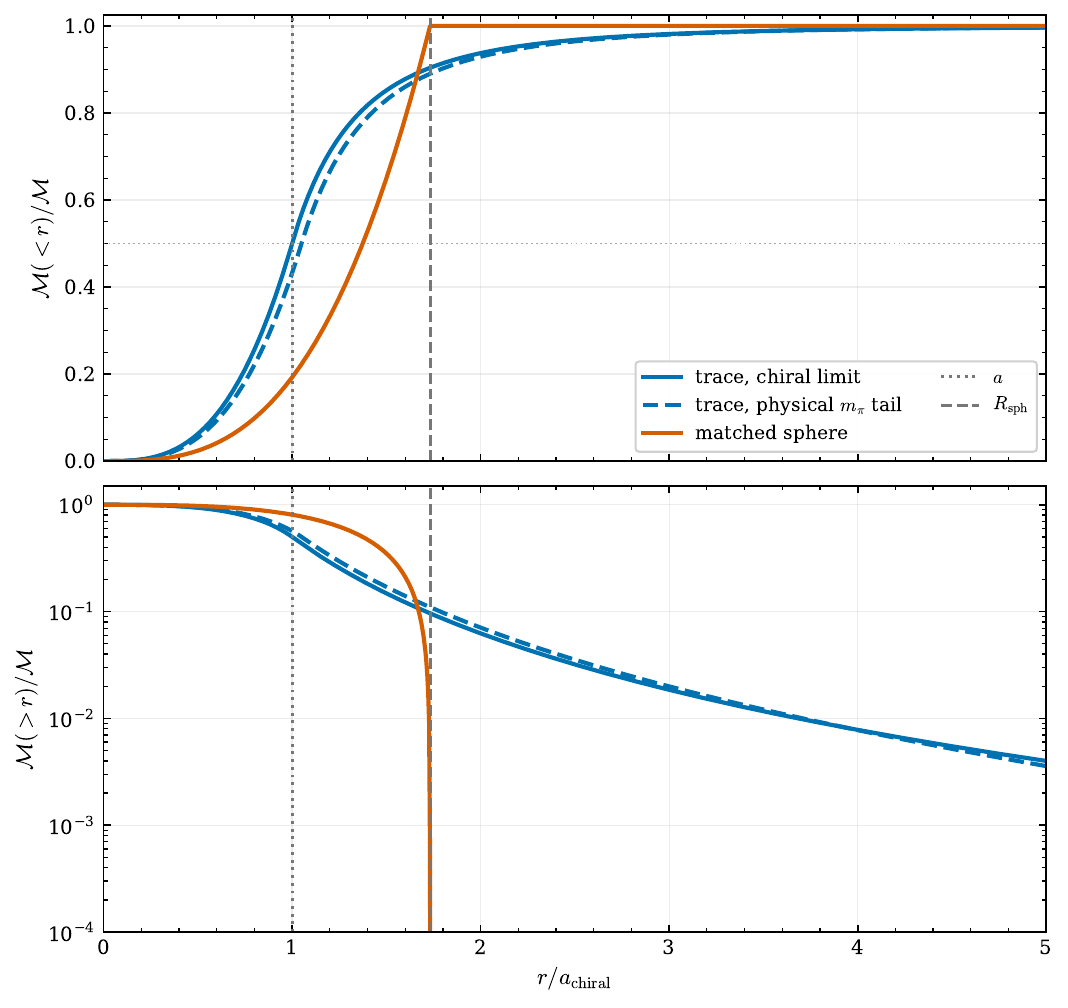}
 \caption{Enclosed mass (\emph{top}) and remaining mass beyond radius
 $r$ (\emph{bottom}) for the chiral-limit trace profile
 Eq.~\eqref{eq:trace-source} (blue solid), the pion-truncated profile
 Eq.~\eqref{eq:trace-source-pi} (blue dashed) at the neutron core radius, and the
 uniform sphere (orange), all at equal mass and RMS radius. The radial axis
 uses the chiral core radius.}
 \label{fig:source}
\end{figure}

Throughout this work we will compare the scalar profiles sourced by Eq.~\eqref{eq:trace-source} with those sourced by  a uniform sphere
of equal total mass and equal root-mean-square (RMS) radius, i.e.
\begin{equation}
   r_{\rm rms}^2= \frac{4\pi}{{\cal M} } \int_0^\infty r^4 \rho_{\rm tr}(r)\, \dd r,
\end{equation}
so that any
difference in screening is a pure shape effect.  The trace-source mass has
\begin{equation}
 \Mcal=4\pi\!\int_0^\infty\! r^2\rho_{\rm tr}(r)\,\dd r
 =\frac{8\pi}{3}\rho_A a^3,
 \quad
 r_{\rm rms}=\frac{3a}{\sqrt5}.
 \label{eq:trace-mass}
\end{equation}
The factor $8\pi/3$ is twice that of a uniform ball of
density $\rho_A$ and radius $a$, as exactly half of the trace-source mass lies outside $a$ (Fig.~\ref{fig:source}). A uniform sphere with equal mass and RMS radius to the trace-source has
$\rs=\sqrt3\,a$ and $\rhos=2\rho_A/(3\sqrt3)$.

For finite pion mass the positive trace source is
$\rho_{\rm tr}=-\tilde T=\varepsilon-3p=2\varepsilon_2+4V_\pi$, where
$\varepsilon_2=F_\pi^2(F'^2+2\sin^2F/r^2)/8$ and
$V_\pi=F_\pi^2m_\pi^2(1-\cos F)/4$. The quartic Skyrme term
cancels in this combination~\cite{Cebulla:2007ei}.
The asymptotic profile is Yukawa-suppressed,
$F\propto(1+m_\pi r)\,e^{-m_\pi r}/r^2$, which cuts off the trace tail
exponentially beyond $r\sim(2m_\pi)^{-1}$. Including the explicit
mass contribution, the small-field trace is proportional to
$F'^2+2F^2/r^2+2m_\pi^2F^2$. Retaining the phenomenological flat
core, we therefore use
\begin{equation}
 \rho_{\rm tr}^{\pi}(r)=\rho_A^{\pi}
 \begin{cases}
  1, & r<a,\\[2pt]
  (a/r)^6\,g(m_\pi r)/g(m_\pi a), & r>a,
 \end{cases}
 \label{eq:trace-source-pi}
\end{equation}
with
\begin{equation}
 g(y)=\left[6+12y+12y^2+8y^3+3y^4\right]e^{-2y}\,.
 \label{eq:anw-tail}
\end{equation}
The amplitude $\rho_A^{\pi}$ is again fixed by requiring the total
mass to equal the nucleon mass.

The truncation introduces a single
new dimensionless parameter, the ratio of the core radius
to the cutoff radius, $2m_\pi a$. Taking
$m_\pi=139.6\,\mathrm{MeV}$, the cutoff radius is
$(2m_\pi)^{-1}\simeq0.7\,\mathrm{fm}$,  and
$2m_\pi a=\PionXNeutron$ ($\PionXProton$) for the neutron (proton)
core radius adopted below. The truncation barely alters the total tail mass, $49.9\%$ ($49.5\%$) of the neutron (proton) mass, against $50\%$
in the chiral limit, but it removes the
far tail. Replacing $g(m_\pi r)/g(m_\pi a)$ by a pure exponential
$e^{-2m_\pi(r-a)}$ and rematching mass and RMS radius changes the
charges by up to about $35\%$, so we take Eq.~\eqref{eq:trace-source-pi} as our baseline for
the physical tail.

For comparison with previous literature, for  the neutron we adopt the Fermi-screening convention of
Refs.~\cite{Fischer:2024eic,Fischer:2023eww}, $\rs=\NeutronSphereRadiusFm\,\mathrm{fm}$,
and for the proton $\rs=\ProtonSphereRadiusFm\,\mathrm{fm}$ following
Ref.~\cite{Brax:2022olf}. This corresponds to considering nucleons as classical lumps of matter with a well-defined matter distribution. Other assumptions have also been invoked such as modelling the matter distribution of nucleons using their wave-function in a quantum mechanical treatment \cite{Jenke:2020obe}. We do not consider this possibility here and restrict ourselves to a traditional approach where the matter distribution of nucleons can be treated classically. 

The matched trace core radius is
$a=\rs/\sqrt3$ in either case. The pion-truncated profile carries
less of its mass at large radius, so holding the comparison sphere
fixed requires a slightly larger core. We solve for the core radius
at which the truncated profile keeps the same RMS radius, giving
$a=\PionCoreNeutronFm\,\mathrm{fm}$ for the neutron and
$\PionCoreProtonFm\,\mathrm{fm}$ for the proton, so that every
comparison in this work is at equal mass and equal RMS radius for
both tail variants. For illustration, setting $a=\BenchmarkCoreRadiusFm\,\mathrm{fm}$ and
$\Mcal=M_N=\BenchmarkCoreMassMeV\,\mathrm{MeV}$ in Eq.~\eqref{eq:trace-mass} fixes
the amplitude by mass normalization,
$\rho_A=3M_N/(8\pi a^3)=\TraceAmplitudeMeVfm\,\mathrm{MeV/fm^3}$.

All charges below depend on the source only through
dimensionless screening parameters. Expressing them in terms of the chameleon coupling
  $\beta$ (or equivalently $M_c$) and the symmetron coupling $M_s$
  requires a specified source mass and core radius.
Where a specific experimental environment is required, such as $\beta$ conversion in
Fig.~\ref{fig:chameleon-charge}, we adopt the qBOUNCE vacuum
residual-gas density
$\rbg=\BenchmarkBackgroundDensity\,\mathrm{kg/m^3}$~\cite{Fischer:2024eic}. The
near-zone symmetron charges are independent of this choice, and the
constraint recomputations of Sec.~\ref{sec:constraints} use each
experiment's own ambient density.

\section{Chameleon screening charges}
\label{sec:chameleon}

For the chameleon of Eq.~\eqref{eq:chameleon-model} the field around a
screened source develops a thin shell. Inside the screening radius
$R_*$, whose value is fixed by the matching below, the field is pinned
at the interior minimum $\phi_c=\phi_{\min}(\rho_{\rm src})$, where $\rho_{\rm src}$ denotes the
core density of the source, $\rho_A$ for the trace profile and $\rhos$
for the sphere. Across the shell the potential term in Eq.~\eqref{eq:eom} is negligible
and the field is driven by the bare source,
\begin{equation}
 \nabla^2\phi\simeq\frac{\rho(r)}{M_c},
 \label{eq:chameleon-shell}
\end{equation}
before relaxing to the ambient value
$\phibg=\phi_{\min}(\rbg)$.  The exterior field carries the monopole
$\phi=\phibg-D/r+\dots$. Integrating Eq.~\eqref{eq:chameleon-shell} once with the flat-core
conditions $\phi(R_*)=\phi_c$, $\phi'(R_*)=0$ gives
$r^2\phi'(r)=M_c^{-1}\int_{R_*}^{r}\rho(s)\,s^2\,\dd s$. A second integration, matched onto the
exterior solution, fixes $R_*$ and the monopole $D$.  The screening charge is
$\Q=D/D_{\rm point}$ with $D_{\rm point}=\Mcal/(4\pi M_c)$.

All source dependence of the screening charge enters through a thin-shell parameter, one per
source model,
\begin{equation}
 \varepsilon_{\rm src}=
 \frac{M_c\left[\phibg-\phi_{\min}(\rho_{\rm src})\right]}
 {\rho_{\rm src}R_{\rm src}^2}\equiv \frac{M_c\left[\phibg-\phi_{\min}(\rho_{\rm src})\right]}
 {6\Phi_{\rm src} \mpl^2}\,,
 \label{eq:thin-shell-parameter}
\end{equation}
with $\Phi_{\rm src}$ the
Newtonian potential of the mass enclosed within $R_{\rm src}$. In what follows, $\varepsilon_{\rm tr}$ denotes Eq.~\eqref{eq:thin-shell-parameter}
evaluated with $(\rho_{\rm src},R_{\rm src})=(\rho_A,a)$ for the trace
source, and $\varepsilon_{\rm sph}$ its evaluation with $(\rhos,\rs)$
for the sphere. The Newtonian potential $\Phi_{\rm src}$ is evaluated at the surface of the source.
We now determine the screening charge for each source model.

\subsection{Uniform sphere}
\label{sec:chameleon-sphere}

For a uniform sphere we follow the classic thin-shell
calculation~\cite{Khoury:2003aq}.  The exterior
solution is the pure monopole $\phi=\phibg-D/r$, and matching
$\phi$ and $\phi'$ at the surface gives
\begin{equation}
 \frac{R_*^2}{\rs^2}=1-2\varepsilon_{\rm sph},
 \qquad
 D=\frac{\rhos}{3M_c}\left(\rs^3-R_*^3\right).
 \label{eq:chameleon-sphere-matching}
\end{equation}
Normalizing by $D_{\rm point}$, the charge is a universal function of
$\varepsilon_{\rm sph}$ alone,
\begin{equation}
\Q_{\rm sph}=
\begin{cases}
1-\left(1-2\varepsilon_{\rm sph}\right)^{3/2},
& \varepsilon_{\rm sph}<\tfrac12,\\[3pt]
1, & \varepsilon_{\rm sph}\ge\tfrac12,
\end{cases}
\label{eq:chameleon-sphere-q}
\end{equation}
whose small-$\varepsilon$ limit recovers the Khoury--Weltman thin-shell
suppression, $\Q_{\rm sph}\simeq3\varepsilon_{\rm sph}=3\,\Delta R/R$~\cite{Khoury:2003aq}.  For the sphere,
Eq.~\eqref{eq:thin-shell-parameter} reduces at leading order in the thin-shell limit to the
thin-shell factor $\Delta R/R$, where  the shell with thickness $\Delta R=\rs-R_*$ contains the only matter that sources the
exterior field.

\subsection{Skyrmion trace source}
\label{sec:chameleon-trace}

For the Skyrmion trace source the tail contributes a particular term to the
exterior solution of Eq.~\eqref{eq:chameleon-shell},
\begin{equation}
 \phi=\phibg-\frac{D}{r}+\frac{\rho_A a^6}{12M_c r^4},
 \label{eq:chameleon-tail-profile}
\end{equation}
and the thin-shell edge can now sit in either the tail or core region.  Matching across the
thin shell gives, for a shell edge inside the core ($R_*<a$),
\begin{equation}
 \frac{R_*^2}{a^2}=\frac32-2\varepsilon_{\rm tr},
 \qquad
 D=\frac{\rho_A}{3M_c}\left(2a^3-R_*^3\right),
 \label{eq:chameleon-core-matching}
\end{equation}
while imposing the flat-core conditions directly on
Eq.~\eqref{eq:chameleon-tail-profile} for a shell edge in the tail
($R_*>a$) gives
\begin{equation}
 \frac{R_*^4}{a^4}=\frac{1}{4\varepsilon_{\rm tr}},
 \qquad
 D=\frac{\rho_A a^6}{3M_c R_*^3}.
 \label{eq:chameleon-tail-matching}
\end{equation}
The same normalization gives
\begin{equation}
\Q_{\rm tr}=
\begin{cases}
\tfrac12\left(4\varepsilon_{\rm tr}\right)^{3/4},
& \varepsilon_{\rm tr}<\tfrac14,\\[3pt]
1-\tfrac12\left(\tfrac32-2\varepsilon_{\rm tr}\right)^{3/2},
& \tfrac14\le\varepsilon_{\rm tr}<\tfrac34,\\[3pt]
1, & \varepsilon_{\rm tr}\ge\tfrac34.
\end{cases}
\label{eq:chameleon-trace-q}
\end{equation}
This form of the solution can be seen to match well with numerically derived screening factors in Fig.~\ref{fig:chameleon-charge}.

\subsection{Interpretation}
\label{sec:chameleon-interpretation}

The chameleon charge has a simple interpretation in terms of Gauss's theorem.  Integrating
Eq.~\eqref{eq:chameleon-shell} over the volume beyond $R_*$ gives
$4\pi r^2\phi'(r)=\Mcal(R_*\!<\!s\!<\!r)/M_c$, so the exterior monopole
is exactly the unscreened mass beyond the screening radius,
\begin{equation}
 \begin{split}
  \Q&=\frac{\Mcal(>R_*)}{\Mcal},\\
  M_c\left[\phibg-\phi_c\right]
  &=\frac{1}{4\pi}\int_{R_*}^{\infty}\!\frac{\Mcal(R_*<s<r)}{r^2}\,\dd r,
 \end{split}
 \label{eq:general-thin-shell}
\end{equation}
with the second relation fixing $R_*$. This gives the lower panel of Fig.~\ref{fig:source}, which shows $\Mcal(>r) / \Mcal$ and can be read as the charge as a function of $R_*$. At higher coupling, corresponding to lower $M_c$ via $M_c=\mpl/\beta$, $R_*$ is pushed outward, so $\Mcal(>R_*)$ and hence the charge are smaller.

The two screened regimes of Eq.~\eqref{eq:chameleon-trace-q}
correspond to the two possible locations of the screening radius: for
weaker
screening ($\tfrac14\le\varepsilon_{\rm tr}<\tfrac34$) $R_*$ lies
inside the core, while for
stronger screening ($\varepsilon_{\rm tr}<\tfrac14$) it recedes into
the $r^{-6}$ tail.  The two expressions meet continuously at $\Q=1/2$,
where $R_*$ crosses the core edge.  In the tail, $R_*$ grows as $\varepsilon_{\rm tr}^{-1/4}$ and the charge
falls only as $\Q_{\rm tr}\propto\varepsilon_{\rm tr}^{3/4}$, whereas, for a uniform sphere,  the
 charge vanishes linearly. For matched constant density sources
$\rhos\rs^2=(2/\sqrt3)\,\rho_Aa^2$,  and since the ambient density lies
far below both source densities, meaning that $\phibg -\phi_{\min}(\rho_{\rm src})\approx \phibg$ in Eq.~\eqref{eq:thin-shell-parameter}, the ratio of the two field excursions
differs from unity by only $\sim10^{-12}$ for the $n=1$ benchmark, so
$\varepsilon_{\rm sph}\simeq(\sqrt3/2)\,\varepsilon_{\rm tr}$ at any
coupling. Deep in screening the charges take their asymptotic forms and
\begin{equation}
 \frac{\Q_{\rm tr}}{\Q_{\rm sph}}\to
 \left(\frac23\right)^{3/2}\varepsilon_{\rm tr}^{-1/4}
 \simeq0.54\,\varepsilon_{\rm tr}^{-1/4},
 \label{eq:chameleon-deep-ratio}
\end{equation}
so the trace source retains parametrically more charge as
$\varepsilon_{\rm tr}\to0$.

\begin{figure}[tbp]
 \centering
 \includegraphics[width=\columnwidth]{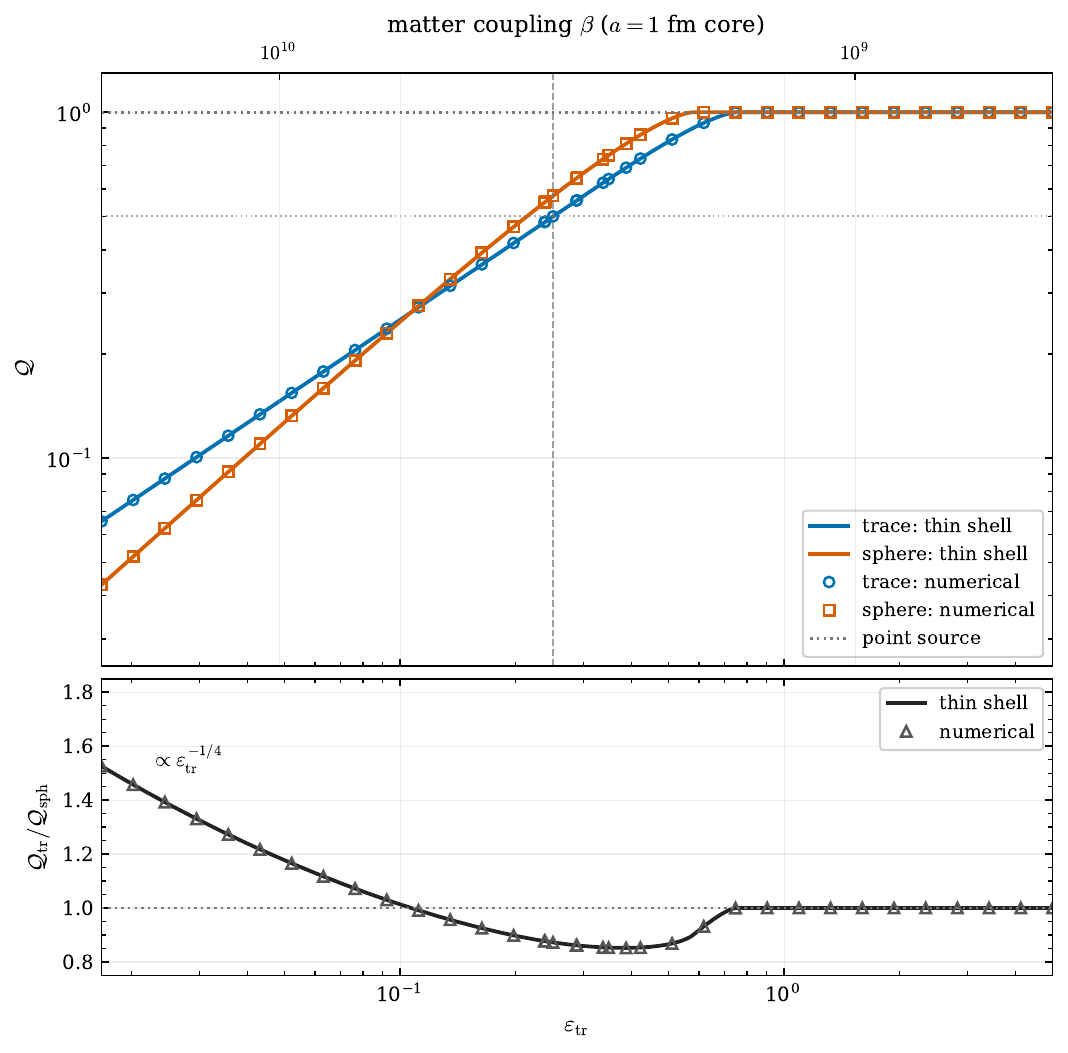}
 \caption{Analytic chameleon charges  against the Skyrmion trace
 thin-shell parameter of Eq.~\eqref{eq:thin-shell-parameter}.  Both
 sources are evaluated at the same physical coupling, for which
 $\varepsilon_{\rm sph}\simeq(\sqrt3/2)\,\varepsilon_{\rm tr}$ (see text). The top axis converts to $\beta$ for the illustrative
 $a=\BenchmarkCoreRadiusFm\,\mathrm{fm}$ core with
 $\Mcal=\BenchmarkCoreMassMeV\,\mathrm{MeV}$ and an equal-mass,
 equal-RMS sphere of radius $\rs=\sqrt3a$.
 The conversion uses $\Lambda=2.4\,\mathrm{meV}$, $n=1$ and
 $\rbg=\BenchmarkBackgroundDensity\,\mathrm{kg/m^3}$.
 \emph{Top:}
 individual charges; open markers are full nonlinear finite-volume
 solutions and the solid curves are the thin-shell results of
 Eqs.~\eqref{eq:chameleon-sphere-q} and \eqref{eq:chameleon-trace-q}.  The
 numerical half-screening point (dashed vertical line) lies at
 $\varepsilon_{\rm tr}=1/4$ within the mesh-refinement estimates
 discussed in the text.
 \emph{Bottom:} the source-shape ratio $\Q_{\rm tr}/\Q_{\rm sph}$, below
 unity near the screening onset, where the sphere is briefly less
 screened, and growing as $(2/3)^{3/2}\,\varepsilon_{\rm tr}^{-1/4}$
 toward strong coupling, where the unscreened tail dominates.}
 \label{fig:chameleon-charge}
\end{figure}

Figure~\ref{fig:chameleon-charge} compares Eqs.~\eqref{eq:chameleon-sphere-q}
and \eqref{eq:chameleon-trace-q} with full
nonlinear finite-volume solutions of Eq.~\eqref{eq:eom} for the model of
Eq.~\eqref{eq:chameleon-model}. The thin-shell charges track the numerics
through the entire screening transition.  At the $n=1$ benchmark the
  numerically determined half-screening point agrees with the thin-shell
  value $\varepsilon_{\rm tr}=1/4$ with a relative discrepancy of $1.4\times10^{-6}$, which decreases
  under mesh refinement.
  This corresponds to $\beta_{1/2}=\ChameleonBetaHalf$ for the illustrative
  $a=\BenchmarkCoreRadiusFm\,\mathrm{fm}$ core  at
  $\Lambda=2.4\,\mathrm{meV}$ and
  $\rbg=\BenchmarkBackgroundDensity\,\mathrm{kg/m^3}$. For $n=1$, fixed mass, $\Lambda$ and $\rbg$, and neglecting
$\phi_{\min}(\rho_{\rm src})$ against $\phibg$ as in
Eq.~\eqref{eq:thin-shell-parameter}, $\varepsilon_{\rm tr}\propto a\beta^{-3/2}$, so
$\beta\propto a^{2/3}$ at fixed charge. The matched neutron core
$a=\ChiralCoreNeutronFm\,\mathrm{fm}$ therefore rescales the top axis
by $\NeutronChameleonBetaRescaling$, giving the thin-shell estimate
$\beta_{1/2}\simeq\NeutronChameleonBetaHalfEstimate$. The matched sphere reaches half screening at
a trace thin-shell parameter smaller by $\ChameleonSphereHalfShift\%$.

\section{Symmetron screening charges}
\label{sec:symmetron}
\subsection{The near-zone behaviour}
For the symmetron of Eq.~\eqref{eq:symmetron-model}, the near zone of
Sec.~\ref{sec:source},
$R\ll r\ll m_{\rm bg}^{-1}$, is very broad throughout the parameter space of interest to laboratory
experiments, since the field's mass at the ambient density,
$m_{\rm bg}=\sqrt{2\left(\mu^2-\rbg/M_s^2\right)}$ in the broken phase,
which reduces to $m_{\rm v}$ in vacuum, is far below the inverse size
of a nucleon. The full numerical equation uses
the total source density, with the residual gas represented by the floor
$\rho_{\rm tot}(r)=\max[\rho_{\rm src}(r),\rbg]$.
Subtracting the background leaves the localized density contrast
$\delta\rho=\rho_{\rm tot}-\rbg$ and the background-potential term
$(m_{\rm bg}^2/2)(\phi^3/\phibg^2-\phi)$.
Neglecting this term in the near zone and taking
$\delta\rho\simeq\rho_{\rm src}$ where the source controls the charge
gives the leading linear problem
\begin{equation}
 \nabla^2\phi=\frac{\rho_{\rm src}(r)}{M_s^2}\,\phi,
 \label{eq:symmetron-near-zone}
\end{equation}
with $\phi\to\phibg$, the ambient value, far beyond the source. The
matter coupling survives as an effective mass
$m_{\rm eff}^2(r)=\rho_{\rm src}(r)/M_s^2$. Far beyond the source the field propagates as if massless, which is what we mean by the
\emph{massless near zone} below. The exterior
near-zone solution is harmonic, $\phi=\phibg-D/r$, and for an unscreened
source, where $\phi\simeq\phibg$ throughout, integrating
Eq.~\eqref{eq:symmetron-near-zone} gives
$D_{\rm point}=\phibg\Mcal/(4\pi M_s^2)$. In the near zone we write
$\Q^{\rm nz}\equiv D/D_{\rm point}$. Because this near-zone equation is
linear in the scalar field, the normalization of $\phi$ drops out, and
$\Q^{\rm nz}$ depends on the source only through the screening parameter
\begin{equation}
 s=\frac{\sqrt{\rho_A}\,a}{M_s}
 \quad\text{(trace)},
 \qquad
 \kappa=\frac{\sqrt{\rhos}\,\rs}{M_s}
 \quad\text{(sphere)}.
\end{equation}

\subsection{Uniform sphere}
\label{sec:symmetron-sphere}

For the uniform sphere the near-zone charge is the classic
symmetry-restored-interior result
\begin{equation}
 \Q_{\rm sph}^{\rm nz}=\frac{3}{\kappa^2}
 \left(1-\frac{\tanh\kappa}{\kappa}\right),
 \label{eq:sphere-near-zone-charge}
\end{equation}
the massless limit of the full Brax--Pitschmann screening factor~\cite{Brax:2017hna,Pitschmann:2020ejb} (Appendix~\ref{app:symmetron}).  The full factor is obtained by matching linearized interior and exterior solutions, and comes in two interior
branches, according to whether the source density restores the
symmetry inside the sphere ($\rhos>\mu^2M_s^2$,
Eq.~\eqref{eq:bp-restored}) or leaves it broken
($\rhos<\mu^2M_s^2$, Eq.~\eqref{eq:bp-broken}). 

\begin{figure}[tbp]
 \centering
 \includegraphics[width=\columnwidth]{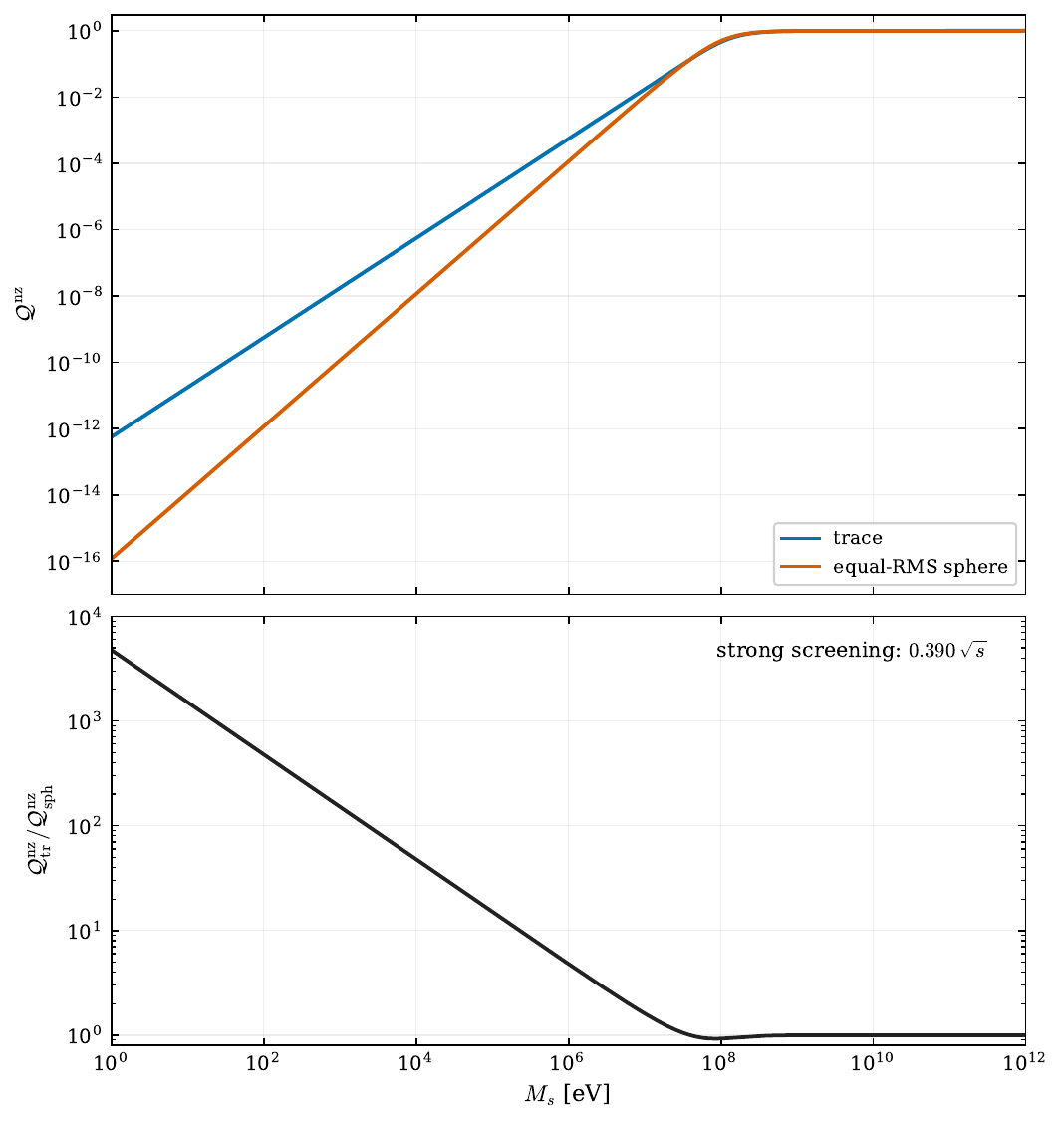}
 \caption{Near-zone symmetron screening and the source-shape enhancement.
 \emph{Top:} the analytical trace charge, Eq.~\eqref{eq:trace-near-zone-charge} (blue),
 and equal-RMS sphere charge, Eq.~\eqref{eq:sphere-near-zone-charge} (orange),
 at fixed mass $\Mcal=\BenchmarkCoreMassMeV\,\mathrm{MeV}$,
 trace core $a=\BenchmarkCoreRadiusFm\,\mathrm{fm}$ and sphere radius
 $\rs=\sqrt3a$.
 \emph{Bottom:} their ratio, with $\kappa=\sqrt{2/\sqrt3}\,s$.
 In strong screening the trace charge scales as $s^{-3/2}$, whereas the
 sphere charge scales as $\kappa^{-2}$, giving the ratio
 $0.390\sqrt{s}$. These near-zone results have no independent dependence
 on $\mu$.}
 \label{fig:symmetron-map}
\end{figure}

\subsection{Skyrmion trace source}
\label{sec:symmetron-trace}

The Skyrmion trace source admits an equally explicit charge.  In the tail the inversion $z=a/r$ with $\chi(z)=\phi(a/z)$ reduces
the radial equation for $r\phi$ to the Weber-type equation
$\chi''=s^2z^2\chi$, solved by $\sqrt z\,I_{\pm1/4}(sz^2/2)$.  Matching to
the evolution in the core, where $r \phi \sim \sinh (sr/a)$, at $z=1$ ($r=a$) gives
(Appendix~\ref{app:symmetron})
\begin{equation}
 \Q_{\rm tr}^{\rm nz}=\frac{3\,\Gamma(3/4)}{\Gamma(1/4)}\,
 \frac{s\,I_{3/4}(s/2)+(s\coth s-1)\,I_{-1/4}(s/2)}
      {s^{5/2}\,[\coth s\,I_{1/4}(s/2)+I_{5/4}(s/2)]}\,.
 \label{eq:trace-near-zone-charge}
\end{equation}
This satisfies $\Q\to1$ as $s\to0$, and deep in screening
\begin{equation}
 \Q_{\rm tr}^{\rm nz}\to\frac{3\,\Gamma(3/4)}{\Gamma(1/4)}\,s^{-3/2}
 \simeq1.014\,s^{-3/2},
\end{equation}
decaying more slowly than the sphere's $3/\kappa^2$ for the same reason as
in the chameleon, namely that the $r^{-6}$ tail is never fully screened and always contributes
to the exterior monopole.

At the illustrative $a=\BenchmarkCoreRadiusFm\,\mathrm{fm}$ core with $\mu=0.05\,\mathrm{eV}$ and $M_s=100\,\mathrm{MeV}$, the charge in Eq.~\eqref{eq:trace-near-zone-charge} agrees with the full nonlinear
boundary-value solution at the $10^{-6}$ level.
Figure~\ref{fig:symmetron-map} shows the source-shape correction in the
near zone. The ratio approaches unity for an unscreened source and grows
as $0.390\sqrt{s}$ toward strong coupling. For a strongly screened trace 
the field responds only once the
local Compton wavelength exceeds the radius, $m_{\rm loc}(r)\,r\simeq1$
with $m_{\rm loc}^2=\rho(r)/M_s^2$. For the $r^{-6}$ tail this is the
\emph{turning radius} $r_t\equiv a\sqrt s$, which sets the exterior
monopole and plays the role of the matching radius $R_*$ of the
chameleon. For the sphere the matching radius is simply $r_t=\rs$. The
near-zone solution assumes $m_{\rm bg}r_t\ll1$, so the separation of
$r_t$ from $m_{\rm bg}^{-1}$, rather than $m_{\rm bg}a$ alone, measures
where the massless treatment applies. Appendix~\ref{app:symmetron-flux}
examines the finite-mass radial flux and the existence of a near-zone
plateau.

\section{Pion-truncated screening charges}
\label{sec:pion}

The chameleon and symmetron charges for a $B=1$ hedgehog Skyrmion can be computed in the same way once the source tail is exponentially  truncated at $r\sim(2m_\pi)^{-1}$, as in Eq.~\eqref{eq:trace-source-pi}. For the chameleon, the two integrals of
Eq.~\eqref{eq:general-thin-shell} apply to any tail. For the truncated
tail they evaluate in terms of incomplete gamma
functions. For the symmetron the massless near-zone problem remains a one-parameter linear equation. Both are checked against full nonlinear solutions
  at the representative parameters specified below.

\begin{figure*}[tbp]
 \centering
 \includegraphics[width=\textwidth]{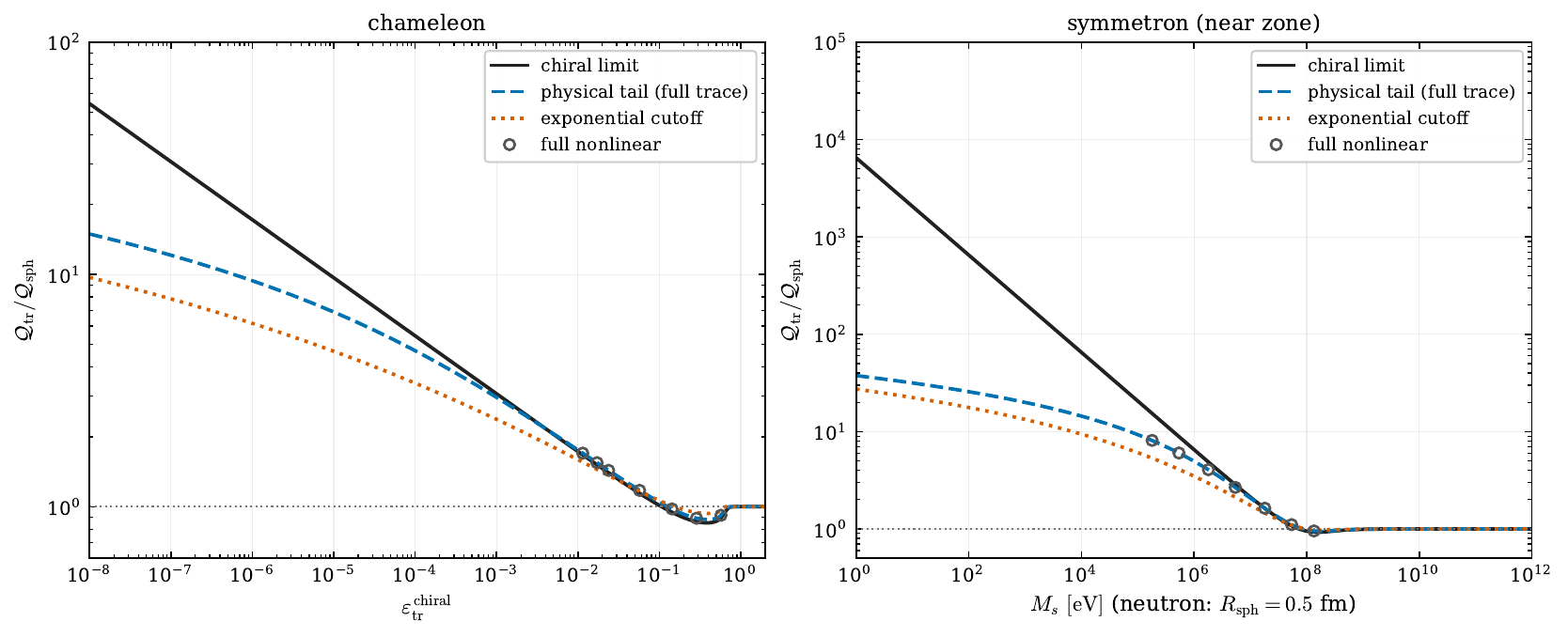}
 \caption{Source-shape charge ratio with the physical pion mass and resulting exponential truncation of the Skyrmion density profile, for the
 chameleon against the trace thin-shell parameter (\emph{left}) and for
 the symmetron in the massless near zone against $M_s$ for the neutron
 core radius (\emph{right}).  Solid black curves are the chiral limit;
 dashed curves use the physical massive-pion tail of
 Eq.~\eqref{eq:trace-source-pi}; dotted curves replace $g$ by a pure
 exponential, with deviations up to about $35\%$ over the plotted ranges.  Open circles are  full nonlinear solutions with the truncated source at  $\PionValidationCount$ points per theory, spanning  $\varepsilon_{\rm tr}^{\rm chiral}=\PionValidationChamMin$ to $\PionValidationChamMax$  and $M_s=\PionValidationSymMassMin$ to $\PionValidationSymMassMax\,\mathrm{eV}$, the latter at $\mu a=10^{-4}$ in units of the truncated core radius and zero ambient density.}
 \label{fig:pion-cutoff}
\end{figure*}

Figure~\ref{fig:pion-cutoff} compares the resulting source-shape ratios
with the chiral limit.  The truncation does not switch the mechanism
off, since an exponential tail still has unscreened mass beyond any screening
radius, so the charge ratio keeps growing logarithmically rather than
as a power law.  Near the
screening onset the truncation has a small effect, since the
tail carries a few percent of the source mass at any coupling. Deep in screening $\Q_{\rm tr}/\Q_{\rm sph}$ is suppressed relative to the chiral limit. The chameleon ratio at
$\varepsilon_{\rm tr}=10^{-5}$ falls from $\PionChamRatioChiral$ to
$\PionChamRatioPhys$, and the neutron symmetron ratio from
$\PionSymRatioChiralLow$ to $\PionSymRatioPhysLow$ at
$M_s=100\,\mathrm{eV}$, from $\PionSymRatioChiralMid$ to
$\PionSymRatioPhysMid$ at $10\,\mathrm{keV}$, and from
$\PionSymRatioChiralHigh$ to $\PionSymRatioPhysHigh$ at
$1\,\mathrm{MeV}$.

\section{Laboratory bounds}
\label{sec:constraints}

The charges derived above determine the response whenever the experimental probe or source is an individual nucleon. For the chameleon neutron probes, the charge varies with the ambient field and must be integrated to obtain the neutron's potential energy. The near-zone symmetron charge is field-independent, and the atomic Yukawa signal is linear in the source charge evaluated at the chamber field. With these prescriptions, exclusion regions can be recomputed without repeating the experimental
analysis.  We have done this for both the chiral-limit and
pion-truncated profiles, for the single-nucleon laboratory probes with
published screened-scalar constraints: qBOUNCE gravitational resonance
spectroscopy~\cite{Cronenberg:2018qxf,Fischer:2024eic}, neutron
interferometry~\cite{Fischer:2023eww,Fischer:2024eic}, and hydrogen
$1s$--$2s$ spectroscopy~\cite{Brax:2022olf}.  Muonium
spectroscopy~\cite{Brax:2022olf} is sourced by a pointlike antimuon, so it is
insensitive to the nucleon model and serves as a fixed
reference. Similarly, constraints from scalar oblique corrections to electroweak precision observables, probed with high-energy particle colliders, do not involve the screening of nucleons and so are unchanged \cite{Brax:2009aw}. Constraints from macroscopic sources (torsion balances, lunar
laser ranging, Casimir experiments) are unaffected, since their screening is
governed by the thin shell of the bulk body, not by nucleon substructure.
Atom-interferometry bounds~\cite{Hamilton:2015zga,Jaffe:2016fsh} sit
between these cases. The source mass is macroscopic, but wherever the
atomic probe's nucleus is screened its charge depends on how the trace
tails of $A$ nucleons combine at the nuclear surface, a multi-nucleon
problem beyond the single-nucleon scope of this work.

\subsection{Experiments and observables}
\label{sec:constraints-observables}

Here we summarize how each experiment works, the observable the
published analysis computes, and the point at which the nucleon charge
enters.  All other inputs follow the published analyses.

\subsubsection{qBOUNCE}
\label{sec:observables-qbounce}

Ultracold neutrons reflect from polished horizontal surfaces at any
angle of incidence, and above a mirror in the Earth's field they occupy
discrete gravitational bound states.  The eigenfunctions are Airy
functions, $\psi_k(z)\propto\mathrm{Ai}(z/z_0-\epsilon_k)$, where
$-\epsilon_k$ is the $k$-th zero of $\mathrm{Ai}$. The characteristic
gravitational length scale of the bounce, for the neutron mass $m_n$, is
$z_0=(2m_n^2g)^{-1/3}\simeq5.9\,\mu\mathrm{m}$, and the energies
$E_k=E_0\epsilon_k$, with $E_0=m_ngz_0\simeq0.60\,\mathrm{peV}$, form
the ladder $\{1.41,2.46,3.32,4.08,\dots\}\,\mathrm{peV}$, set by
gravity alone.

A scalar field sourced by the mirror adds the adiabatic potential
\begin{equation}
 \delta V(z)=m_n\int_{\phi_{\rm ref}}^{\phi_{\rm mirror}(z)}
 \Q_n(\varphi)A_{,\varphi}(\varphi)\,\dd\varphi,
 \label{eq:screened-potential}
\end{equation}
where $\phi_{\rm mirror}(z)$ is the static profile above the mirror
and $\Q_n (\varphi)$ is the neutron screening charge of Secs.~\ref{sec:chameleon}--\ref{sec:symmetron} evaluated in the local mirror-generated field. The arbitrary reference field changes only a constant energy offset.
For the chameleon, $A_{,\phi}=1/M_c$ at leading order and we integrate
the local thin-shell charge.  For the symmetron, the field-independent
near-zone charge gives $\delta V(z)=\Q_n m_n\bigl[\phi_{\rm mirror}^2(z)-\phi_{\rm ref}^2\bigr]/(2M_s^2)$.
At first order the level shifts are the expectation values
$\delta E_k=\langle\psi_k|\delta V|\psi_k\rangle$, which suffices for
the chameleon. The symmetron layer can be deep enough that we instead
solve for the full perturbed spectrum, following the published
nonperturbative treatment (Sec.~\ref{sec:constraints-symmetron}).  The
qBOUNCE experiment at the Institut
Laue--Langevin~\cite{Cronenberg:2018qxf} drives the $1\to3$ and
$1\to4$ transitions mechanically and measures
$E_{13}=(1.9222\pm0.0054)\,\mathrm{peV}$ and
$E_{14}=(2.6874\pm0.0074)\,\mathrm{peV}$, a precision of a few
$10^{-15}\,\mathrm{eV}$.

\subsubsection{Neutron interferometry}
\label{sec:observables-interferometry}

Neutrons are affected by the presence of the scalar field and interfere when a neutron beam is split into two separate paths that recombine after each crosses a chamber. A neutron of kinetic
energy $E_{\rm kin}=k^2/(2m_n)$, where $k$ is the neutron wavevector,
crossing a region of potential $\delta V$ propagates with the local
wavevector $k(\ell)=k\sqrt{1-\delta V(\ell)/E_{\rm kin}}$ at position
$\ell$ along its flight path, the index-of-refraction relation of
neutron optics~\cite{RauchWerner:2015}.  The phase accumulated relative to
the empty cell,
$\delta\Phi=\int_{\rm path}\left[k(\ell)-k\right]\dd\ell$, is at
leading order in $\delta V/E_{\rm kin}$
\begin{equation}
\label{eq:interferometry-phase}
\delta \Phi= -\frac{m_n}{k}
\int_{\rm path} \dd\ell\,\delta V(\phi_{\rm chamber})\,,
\end{equation}
where $\phi_{\rm chamber}$ is the field profile in the
chamber~\cite{Lemmel:2015kwa}, and $\delta V$ is the same adiabatic
energy of Eq.~\eqref{eq:screened-potential}, evaluated at the chamber
field with a common reference for both paths.

The expansion requires $|\delta V|\ll E_{\rm kin}\simeq11\,\mathrm{meV}$ for the $2.72$\,\AA\ beam. This is stronger than the condition $\delta V<E_{\rm kin}$ for a real local wavevector. Larger transmitting potentials require the unexpanded expression for $k(\ell)$, and near a turning point, reflection must also be treated.

The searches of
Refs.~\cite{Lemmel:2015kwa,Fischer:2023eww} at the S18 instrument of
the Institut Laue--Langevin send neutrons of wavelength $2.72$\,\AA\
through two chambers in the same housing: a vacuum cell of length
$L=9.4\,\mathrm{cm}$, filled with helium through a leak valve, and a
reference chamber open to ambient air. Inside either one the scalar
is suppressed at the dense walls, so $\delta\Phi$ depends on the
transverse position at which the beam crosses. Because
only phase differences are measurable, two differential observables
isolate the in-cell field.  We resolve the field in both the radial and longitudinal directions
of the finite cylinder, including its end walls, and integrate
$\delta V(\phi)$ along each beam path before forming the differences.
The effective length of $8.4\,\mathrm{cm}$ used in
  Ref.~\cite{Lemmel:2015kwa} was obtained by integrating $\phi$
  for an unscreened neutron in the $n=1$ chameleon model. The screened energy is nonlinear in $\phi$, so the
  same effective length need not apply.

In \emph{profile mode} the chamber assembly is translated sideways, so
the beams cross their chambers on the axis and $1.5\,\mathrm{cm}$
off-axis, with the cell at $10^{-4}\,\mathrm{mbar}$,
\begin{align}
 \alpha&=\left[\delta\Phi_{\rm vac}(0)-\delta\Phi_{\rm vac}(1.5\,\mathrm{cm})\right]
 \nonumber\\
 &\quad-\left[\delta\Phi_{\rm air}(0)-\delta\Phi_{\rm air}(1.5\,\mathrm{cm})\right]
 =0.56^\circ\pm2.50^\circ.
 \label{eq:interferometry-alpha}
\end{align}
In \emph{pressure mode} both beams stay on axis and the helium
filling is stepped between $P_0=2\times10^{-4}$ and
$P_1=10^{-2}\,\mathrm{mbar}$,
\begin{equation}
 \gamma=\delta\Phi_{P_0}(0)-\delta\Phi_{P_1}(0)
 =0.37^\circ\pm3.53^\circ.
 \label{eq:interferometry-gamma}
\end{equation}
Each scalar phase contribution is negative, but the observables are
differences between two cells, so their signs depend on which cell
hosts the larger contrast. The pressure mode is always negative. The
profile mode is negative wherever the vacuum cell carries more
transverse contrast than the air reference, but a nearly flat vacuum
profile can carry less, in which case $\alpha$ is positive. This occurs for a very narrow 
mass window at $1\,\mathrm{meV}$. The one-sided $95\%$ thresholds are
therefore chosen by the sign of the predicted phase:
$\alpha<-3.55^\circ$ or $\alpha>+4.67^\circ$, and
$\gamma<-5.44^\circ$ or $\gamma>+6.18^\circ$.

The in-cell profiles are computed per model. The symmetron field vanishes at the side and end walls and is nonzero only above the finite-cylinder threshold $m_{\rm bg}R_{\rm c}\simeq3.53$, with $R_{\rm c}=2\,\mathrm{cm}$ and $m_{\rm bg}$ evaluated at the chamber gas density. The chameleon fills the cell with the $n=1$ bubble profile used in Sec.~\ref{sec:constraints-chameleon}.

\subsubsection{Hydrogen and muonium}
\label{sec:observables-atoms}

The analysis of Ref.~\cite{Brax:2022olf} uses the hydrogen $1s$--$2s$
spectroscopy measurement of Ref.~\cite{Parthey:2011lfa}, accurate to
four parts in $10^{15}$, which caps any scalar-induced shift at
$4.3\times10^{-14}\,\mathrm{eV}$, and the muonium measurement of
Ref.~\cite{Meyer:1999cx}, six orders of magnitude less precise at
$4.1\times10^{-8}\,\mathrm{eV}$.  The screened nucleus adds a Yukawa
perturbation on top of the ambient chamber field,
\begin{equation}
 \delta V(r)=-\Q_p\,\frac{m_pm_e}{4\pi M_c^2}\,
 \frac{e^{-m_{\rm out}r}}{r},
 \label{eq:atoms-yukawa}
\end{equation}
with $M_c^{-2}\to\phi_{\rm out}^2/M_s^4$ for the symmetron, where
$m_{\rm out}$ and $\phi_{\rm out}$ are the scalar mass and field value
in the vacuum chamber. For the chameleon, $\phi_{\rm out}$ is the smaller of the minimum of the effective potential at the
chamber gas density (the chamber minimum), and the wall-limited value,
with effective chamber radii of $1\,\mathrm{cm}$ for the hydrogen
apparatus and $0.8\,\mathrm{cm}$ for muonium, and residual-gas
densities of $10^{-14}$ and $10^{-13}\,\mathrm{g\,cm^{-3}}$
respectively~\cite{Brax:2022olf}. The wall-limited value is the field at which the chameleon's
in-chamber Compton wavelength matches the cavity size,
$m(\phi_{\rm vac})\simeq1/R$, giving
$\phi_{\rm vac}=\xi\,(2\Lambda^5R^2)^{1/3}$ with $\xi=0.55$ for a
spherical chamber, so that $m(\phi_{\rm vac})\,R=\xi^{-3/2}\simeq2.4$.
We take $\phi_{\rm out}=\min(\phi_{\min},\phi_{\rm vac})$, so the field
reaches the chamber minimum where $m(\phi_{\min})\,R>\xi^{-3/2}$, and
otherwise stops at $\phi_{\rm vac}$.

First-order perturbation theory in the hydrogenic states gives the
closed expectation values
\begin{align}
 \Big\langle1s\Big|\frac{e^{-mr}}{r}\Big|1s\Big\rangle
 &=\frac{4}{a_0(2+ma_0)^2},
 \nonumber\\
 \Big\langle2s\Big|\frac{e^{-mr}}{r}\Big|2s\Big\rangle
 &=\frac{1+2m^2a_0^2}{4a_0(1+ma_0)^4},
 \label{eq:atoms-matrix-elements}
\end{align}
with $a_0$ the Bohr radius and $m=m_{\rm out}$, and the observable is
the differential shift $|\delta E_{2s}-\delta E_{1s}|$, exactly linear
in  $\Q_p$, which is evaluated at the ambient field
of the apparatus.  The muonium source is the point-like anti-muon, taken
to be unscreened with $\Q=1$ everywhere in parameter space.

\subsubsection{Exclusion criteria and validity cuts}

A parameter point is excluded when the predicted signal leaves the
$2\sigma$ window (qBOUNCE), crosses the one-sided $95\%$ threshold
(interferometry), or exceeds the spectroscopic cap (hydrogen and
muonium). Table~\ref{tab:cuts} summarises the validity and display
cuts that delimit every constraint region in the following sections.

\begin{table*}[tbp]
 \caption{Validity and display cuts applied to the regions of
 Figs.~\ref{fig:constraints-chameleon} and
 \ref{fig:constraints-symmetron}.  Each cut bounds the domain of an
 approximation or of the comparison with the published contours. $A-1$ is
 $\phi/M_c$ for the chameleon and $\phi^2/(2M_s^2)$ for the
 symmetron, evaluated at the ambient field of each apparatus.
 $E_{\rm kin}$ is the kinetic energy of the interferometer beam and
 $V$ the in-cell potential of the screened neutron,
 Eq.~\eqref{eq:screened-potential} from a zero reference field.}
 \label{tab:cuts}
 \begin{ruledtabular}
 \begin{tabular}{p{2.8cm}p{3.0cm}p{10.6cm}}
 Condition & Applies to & Criterion \\
 \hline
 Weak coupling & all chameleon regions & $A-1\le0.1$ \\
 Weak coupling & all symmetron regions & $A-1\le0.02$, the termination read off the published contours~\cite{Fischer:2024eic} \\
 Screening range & all regions & chameleon $m_{\rm out}R_*\le1$, symmetron $m_{\rm bg}r_t\le1$, at the field that generates the signal; removes only strong-coupling points of the chiral chameleon hydrogen and interferometry regions, where the boundary is not drawn (Sec.~\ref{sec:constraints-chameleon}) \\
 Linear phase & interferometry & $V\le0.1\,E_{\rm kin}$ for Eq.~\eqref{eq:interferometry-phase}; symmetron points with $0.1E_{\rm kin}<V<E_{\rm kin}$ and $V\ge E_{\rm kin}$ are shown as beyond first order and transmission unverified (Sec.~\ref{sec:constraints-symmetron}) \\
 Wall equilibration & interferometry, chameleon & Compton wavelength at the wall density below $0.25\,\mathrm{mm}$,  so that the field reaches its wall minimum inside the aluminium walls \\
 Interaction range & qBOUNCE & vacuum Compton wavelength below $1\,\mathrm{mm}$~\cite{Fischer:2024eic} \\
 Broken phase & all symmetron regions & $\mu^2>\rbg/M_s^2$ and the chamber large enough to support the vev (finite-cylinder threshold $m_{\rm bg}R_{\rm c}\simeq3.53$ for the interferometer, $\mu^{-1}<R$ for spectroscopy); symmetric-phase points carry no signal \\
 \end{tabular}
 \end{ruledtabular}
\end{table*}

\subsection{Chameleon bounds}
\label{sec:constraints-chameleon}

\begin{figure*}[tbp]
 \centering
 \includegraphics[width=0.72\textwidth]{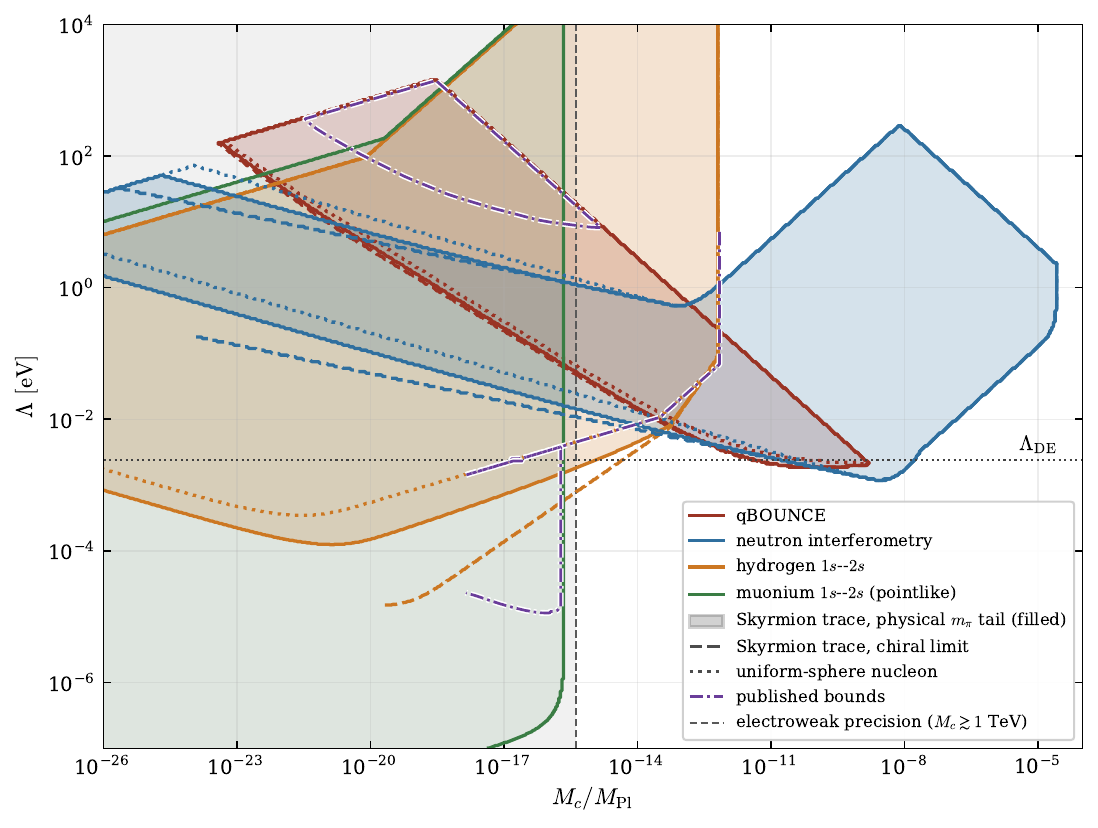}
 \caption{Combined chameleon ($n=1$) exclusion regions from qBOUNCE (red), neutron interferometry (blue), and
 hydrogen (orange) and muonium (green) $1s$--$2s$ spectroscopy, following the analyses of
 Refs.~\cite{Fischer:2024eic,Brax:2022olf}.  
Filled regions with a solid boundary use the Skyrmion trace source with
the pion-truncated profile of Eq.~\eqref{eq:trace-source-pi}, the
physical nucleon model; dashed boundaries use the equal-mass, equal-RMS
Skyrmion trace source in the chiral limit; dotted boundaries use the
uniform-sphere nucleon. All three use the same thin-shell charge.
 Muonium (solid green line) is point-like-sourced and fixed, as is the (grey dashed) bound from electroweak precision tests.  All three hydrogen bounds cross the dark-energy scale
 $\Lambda_{\rm DE}=2.4\,\mathrm{meV}$ at strong coupling. The
 chiral-limit trace proton reaches furthest below it. At strong
 coupling the chiral hydrogen region, and the lower edge of the chiral
 interferometry band, end on the thin-shell validity line
 $m_{\rm out}R_*=1$ (Sec.~\ref{sec:constraints-chameleon}). This edge
 is not drawn. The dashed boundaries stop where the matching ceases to be valid, not where the signal disappears. Purple dot-dash lines are previously published constraints from Refs.~\cite{Fischer:2024eic,Brax:2022olf} which assume a uniform sphere nucleon.}
 \label{fig:constraints-chameleon}
\end{figure*}

Figure~\ref{fig:constraints-chameleon} shows the chameleon
constraints for $n=1$ in the $(M_c/\mpl,\Lambda)$ plane, with
previously published regions also displayed. Throughout,
$\Lambda_{\rm DE}=2.4\,\mathrm{meV}$ denotes the dark-energy scale.
The chameleon regions are restricted to weak coupling,
$A(\phi)-1\le0.1$, so that the higher-order matter-coupling terms
dropped by every treatment stay small, and to the domain where the
thin-shell matching is self-consistent at the ambient field the probe
samples. The matching needs $R_*$ well inside the local Compton
wavelength, and we adopt the unit ratio $m_{\rm out}R_*\le1$. Only the
chiral regions of hydrogen and interferometry lose parameter space to this rule,
and there the boundary is not drawn along the removed edge. The
interferometry region additionally requires the in-cell potential of
the screened neutron, the same charge integral that enters the phase,
to stay below a tenth of the beam's kinetic energy, which is the
validity domain of Eq.~\eqref{eq:interferometry-phase}, and keeps the
Compton wavelength at the wall density below $0.25\,\mathrm{mm}$ so
that the field settles to its bubble profile inside the cell.
Following Ref.~\cite{Fischer:2024eic}, the qBOUNCE regions keep the
vacuum interaction range below $1\,\mathrm{mm}$, so that the vacuum
chamber can be neglected. All other inputs follow the published
analyses.

\paragraph{qBOUNCE.}
For the uniform sphere neutron the exclusion covers
$\ConstraintQbounceMassMin\lesssim M_c/\mpl\lesssim\ConstraintQbounceMassMax$, reaching from $\Lambda\simeq\ConstraintQbounceLambdaMax\,\mathrm{eV}$ down across the
dark-energy scale at its weak-coupling tip. The bounds differ from
the published $n=1$ region of Ref.~\cite{Fischer:2024eic}, in both
the charge and energy prescriptions. That analysis screens the neutron
with a linear-response factor built on a constant interior chameleon
mass and uses a fixed-vacuum-charge potential. Here the charge follows
from the nonlinear thin-shell matching of
Eq.~\eqref{eq:chameleon-sphere-q} and is integrated over the local
field as in Eq.~\eqref{eq:screened-potential}. A reference calculation
using the previously published methods reproduces its region, including
the curved lower boundary.

The Skyrmion changes the uniform sphere result at the strong-coupling edge.  The edge exists because the predicted energy shift is proportional to $(1/M_c)\int^\phi\Q\,\dd\varphi$. Toward smaller $M_c$ the explicit $1/M_c$, the coupling of
an unscreened nucleon, grows, but the neutron's charge
falls faster ($\Q_{\rm sph}\simeq3\,\varepsilon_{\rm sph}$ with
$\varepsilon_{\rm sph}\propto M_c^{3/2}$), so the shift falls back
inside the measured windows and sensitivity is lost. The trace
neutron delays this, as its charge falls as
$\Q_{\rm tr}\propto\varepsilon_{\rm tr}^{3/4}$ (Sec.~\ref{sec:chameleon}), which widens the exclusion by a factor $\ConstraintQbounceChiralGain$ in the matter coupling in the chiral limit at $\Lambda_{\rm DE}$, and $\ConstraintQbouncePionGain$ with the physical pion mass.

Truncation removes the
tail beyond
the cutoff radius $(2m_\pi)^{-1}\simeq2.3\,a$, and the enhancement is reduced to logarithmic growth only once the screening radius passes that cutoff. At $\Lambda_{\rm DE}$ the screening radius at the vacuum field is $R_*\simeq5.9$ truncated core radii, well past the cutoff, yet most of the chiral-limit gain remains.

\paragraph{Neutron interferometry.}
The original chameleon analysis of this experiment,
Ref.~\cite{Lemmel:2015kwa}, solved the field in the actual square cell
and fit the full multi-position phase data, but treated the neutron as
an unscreened test particle, an approximation that holds near its
weak-coupling bound and fails toward stronger coupling, where the
neutron screens.  Ref.~\cite{Fischer:2024eic} revisited the experiment
with source screening included, bounded the largest attainable phase
using its linear-response charge, and concluded that a more detailed
analysis would be ``futile''.  This conclusion depends on the choice of charge equation. The chameleon mass inside the neutron is always
very large, so a charge built on it is negligible, whereas the
thin-shell criterion leaves the neutron barely screened wherever
$\varepsilon_{\rm tr}$ is of order unity.

With the thin-shell charge
the recomputed exclusion, using both interferometry modes with the
$n=1$ bubble profile in the cell, has no strong-coupling
termination. It spans more than twenty-one decades of matter
coupling, from $M_c/\mpl\simeq\ConstraintInterferometryMassMax$, where the
weak-coupling sensitivity edge meets the wall-equilibration validity
cut, down past the strong-coupling edge of our computed grid at
$M_c/\mpl=10^{-26}$.  The band runs from 
$\Lambda\simeq\ConstraintInterferometryLambdaMin \,\mathrm{eV}$ up to
$\Lambda\simeq\ConstraintInterferometryLambdaMax\,\mathrm{eV}$.  Its lower boundary is the
sensitivity threshold, while its ceiling is the validity domain of
Eq.~\eqref{eq:interferometry-phase} evaluated on the screened
potential, with the $A-1\le0.1$ cut taking over at the extreme edge
of the grid.  We do not repeat the square-cell, multi-position treatment of Ref.~\cite{Lemmel:2015kwa}, but keep the cylindrical two-point reduction of the symmetron analysis so that both scalars share one experimental model.  As a result, at the weak-coupling edge, where no
screening physics enters, our bound at $\Lambda=\Lambda_{\rm DE}$ is a factor
of ${\sim}3$ more conservative than that of Ref.~\cite{Lemmel:2015kwa}.

The Skyrmion gain appears on the band's sensitivity edge and
grows with the depth of screening. At fixed coupling the trace
neutron excludes lower $\Lambda$ than the sphere, by $\ConstraintInterferometryChiralGainSixteen$ decades
at $M_c/\mpl=10^{-16}$ rising to $ \ConstraintInterferometryChiralGainTwentyFour$ decades at $10^{-24}$ in the
chiral limit ($\ConstraintInterferometryPionGainSixteen$ and $\ConstraintInterferometryPionGainTwentyFour$ decades, respectively, with the physical tail). On the validity ceiling the ordering reverses, because the larger surviving trace charge reaches the linear-phase limit $V\le0.1E_{\rm kin}$ sooner, so the sphere band extends to higher $\Lambda$ there. This edge, like the strong-coupling end of the chiral band's lower edge, marks the limit of the validity domain and not of the sensitivity. At the latter the thin-shell rule
$m_{\rm out}R_*\le1$ of Table~\ref{tab:cuts}, evaluated at the field
of the cell that generates each phase signal, removes the chiral
neutron's exclusion below $M_c/\mpl\simeq10^{-24}$. The dashed
boundary is left undrawn there.  The sphere and pion-tail neutrons
satisfy the rule throughout their bands.

Using the thin-shell charge restores a neutron-interferometry
exclusion band that is absent with the linear-response prescription, for which the maximal-phase envelope of Ref.~\cite{Fischer:2024eic} remains below the experimental threshold. This region is already excluded by atom interferometry, so the result is an independent constraint and does not extend the combined exclusion.

\paragraph{Hydrogen and muonium.}
 
The sphere and both Skyrmion trace protons use the same thin-shell matching. For comparison, a reference calculation using the published interpolation $\min(3\varepsilon_{\rm sph},1)$, with the interior field neglected, reproduces the hydrogen exclusion of Ref.~\cite{Brax:2022olf}. The right edge, $M_c/\mpl\lesssim\ConstraintHydrogenMassMax$, matches to within $0.05$ decades, and the low-$\Lambda$ boundary to a few hundredths of a decade. Using Eq.~\eqref{eq:chameleon-sphere-q} for the sphere slightly weakens the crossover corner, while leaving the overall coupling reach and minimum excluded $\Lambda$ unchanged on our grid. This boundary is set by screening. Toward small $\Lambda$ the
shrinking chamber field deepens the proton's screening, and the
sphere bound ends near $\Lambda\sim\ConstraintHydrogenSphereLambdaMin\,\mathrm{eV}$, just below the dark-energy scale. The trace
proton delays this cutoff, as for qBOUNCE.  Under deep
screening the sphere charge vanishes linearly in the screening
parameter, whereas the tail charge falls only as its $3/4$ power, so
the exclusion remains to $\Lambda\simeq\ConstraintHydrogenPionLambdaMin\,\mathrm{eV}$
with the physical tail and to
$\Lambda\simeq\ConstraintHydrogenChiralLambdaMin\,\mathrm{eV}$ in the
chiral limit, more than an order of magnitude below the sphere's floor.

Toward strong coupling the chiral screening radius grows without
bound, $R_*=a\,(4\varepsilon_{\rm tr})^{-1/4}$, and the matching of
Sec.~\ref{sec:chameleon} needs a near zone between the shell and the
Yukawa exterior. In this matching
$m_{\rm out}R_*\simeq\sqrt8\,\phi_{\min}(\rho(R_*))/\phi_{\rm out}$,
so the ratio compares the tail density at the shell edge with
$\rho_{\rm eq}=M_c\Lambda^5/\phi_{\rm out}^2$, the density whose
minimum is the chamber field, here the gas density $\rbg$. 
With $m_{\rm out}R_*=1$ the tail at the shell edge is still eight times denser than the gas, corresponding to $R_*\simeq0.5\,a_0$. The electron's orbit is no longer outside the shell, and the
picture of an electron feeling a Yukawa tail from a screened proton
breaks down. We therefore terminate the chiral constraint at this point, where the chiral floor quoted above lies, and Fig.~\ref{fig:constraints-chameleon} does not draw the boundary along it. The sphere's shell never leaves the
source and the pion-truncated screening radius grows only
logarithmically, so these constraints are unaffected. 

Along $\Lambda=\Lambda_{\rm DE}$ the sphere proton excludes
$M_c/\mpl\lesssim\ConstraintHydrogenDEMassSphere$, the pion-tail
Skyrmion $\ConstraintHydrogenDEMassPion$ and the chiral Skyrmion
$\ConstraintHydrogenDEMassChiral$, so at the dark-energy scale the
Skyrmion proton carries the hydrogen bound beyond muonium's
$\ConstraintHydrogenDEMassMuonium$. Part of the newly excluded region was previously
covered only by muonium.  Muonium
itself, sourced by the pointlike antimuon, has no charge to modify. Its bound, $M_c/\mpl\lesssim\ConstraintMuoniumMassMax$, is the same in all
three source models. The published region stops at
$\Lambda\simeq1.2\times10^{-5}\,\mathrm{eV}$, where the in-chamber
mass still satisfies $m_{\rm out}a_0\approx0.02$, a Compton
wavelength of roughly fifty Bohr radii. Ref.~\cite{Brax:2022olf}
evaluates the shift in the light-field limit and truncates the bound
well within that approximation's domain of validity.  The exact
matrix elements of Eq.~\eqref{eq:atoms-matrix-elements} hold at any mass and keep the exclusion at the weak-coupling edge down to
$\Lambda\simeq\ConstraintMuoniumWeakEdgeLambdaMin\,\mathrm{eV}$,
where the finite-range suppression becomes relevant. The global
low-$\Lambda$ extent reaches the lower edge of our scan.

\subsection{Symmetron bounds}
\label{sec:constraints-symmetron}

\begin{figure*}[tbp]
 \centering
 \includegraphics[width=\textwidth]{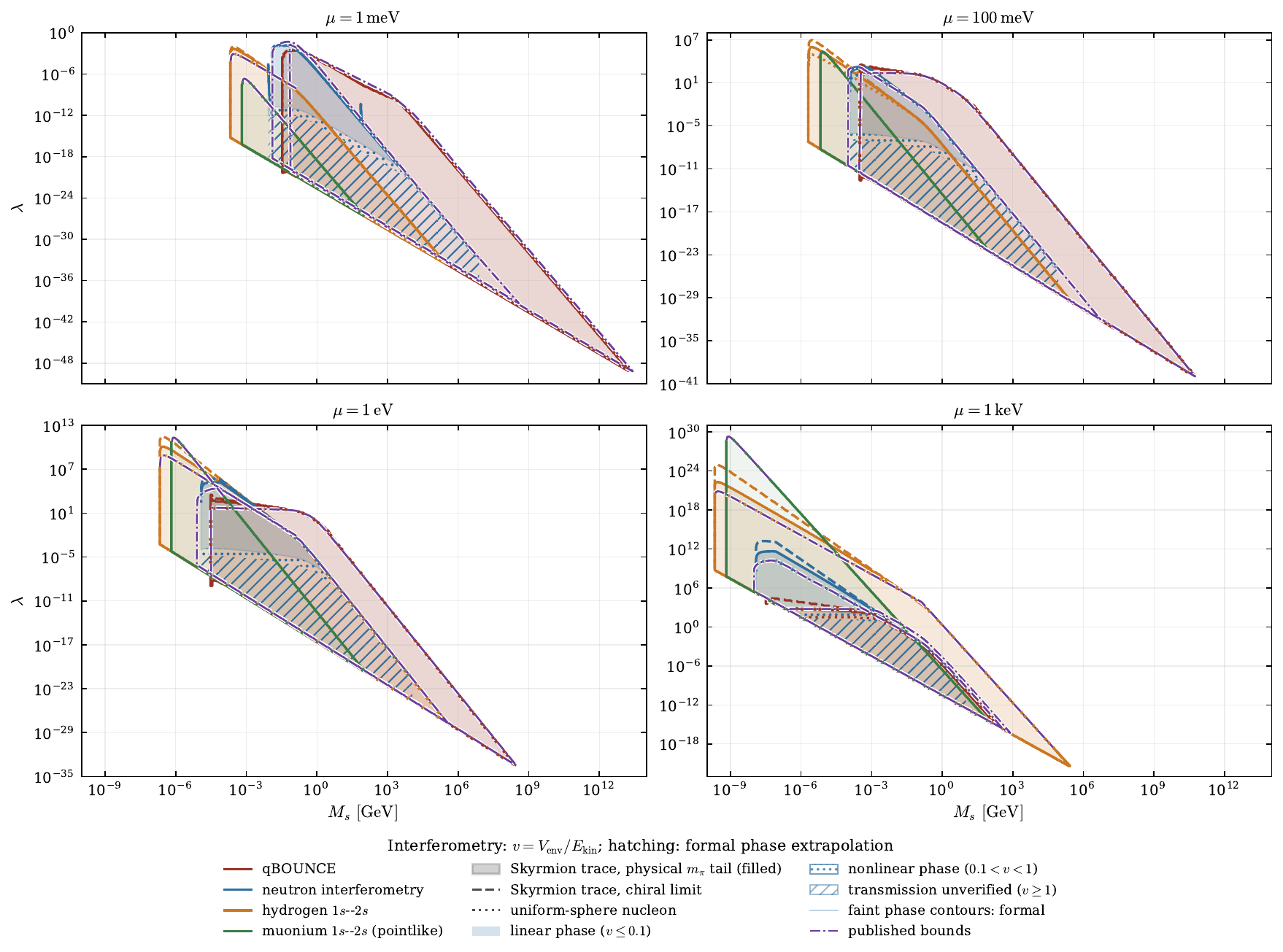}
 \caption{Combined symmetron exclusion regions in fixed-$\mu$ slices from
 qBOUNCE, neutron interferometry, and hydrogen and muonium spectroscopy.
 Line conventions as in Fig.~\ref{fig:constraints-chameleon}. For the filled (physical $m_\pi$ Skyrmion) interferometry region, the shading distinguishes three regimes of the potential
 envelope $v=V_{\rm env}/E_{\rm kin}$: solid shading for $v\le0.1$,
 dots for $0.1<v<1$ (nonlinear phase required), and diagonal hatching
 for $v\ge1$ (transmission unverified). Both hatched regimes show linear-phase extrapolations.
  Faint contour segments fail the corresponding source's
  $v\le0.1$ cut. The published hydrogen and muonium regions of
  Ref.~\cite{Brax:2022olf} apply no $A-1$ cut, and they are shown only
  where $A-1\le0.02$, the display floor used for our regions. The $\mu=1\,\mathrm{keV}$ qBOUNCE regions are conditional on treating the
mirror as a hard wall and on reading the measured transitions as
transitions out of the lowest eigenvalue (see
Appendix~\ref{app:qbounce-states} for discussion).}
 \label{fig:constraints-symmetron}
\end{figure*}

Figure~\ref{fig:constraints-symmetron} shows the corresponding symmetron constraints, with previously published regions also shown. The first-order interferometry phase and spectroscopy observables are
linear in $\Q/\lambda$, so their exclusion boundaries in $\lambda$ are
rescaled by the nucleon's charge ratio. The
qBOUNCE interaction-range cut is that of
Table~\ref{tab:cuts}. We terminate the displayed regions at $A-1\simeq0.02$, the value read off the published contours of Ref.~\cite{Fischer:2024eic} (their stated criterion is $A-1\ll1$), common to every $\mu$ slice and to both neutron experiments. Throughout, the Skyrmion trace charges are evaluated in the massless
near zone, Eq.~\eqref{eq:trace-near-zone-charge} and its
pion-truncated form, while the sphere uses the
full Brax--Pitschmann factor with both interior branches.
The
near-zone treatment is self-consistent wherever its matching radius
$r_t$ of Sec.~\ref{sec:symmetron-trace} lies inside the ambient Compton
wavelength, $m_{\rm bg}r_t\le1$, the symmetron form of the
screening-range rule of Table~\ref{tab:cuts}. This is the case for each $\mu$ slice.
Figure~\ref{fig:symmetron-map} shows the near-zone source-shape enhancement. Full nonlinear source-profile checks support the near-zone approximation
at the few-percent level over the sampled constraint range, including
$\mu=1\,\mathrm{keV}$ down to $M_s=1\,\mathrm{eV}$.
Appendix~\ref{app:symmetron-flux} gives the quantitative comparison and
examines the radial flux at finite background mass.

\paragraph{qBOUNCE.}
We follow the
nonperturbative treatment of Ref.~\cite{Fischer:2024eic}, using the
exact one-mirror profile with its residual-gas density, the perturbed
spectrum from the full Schr\"odinger problem, and the published level
convention, in which the measured transitions are read off the
eigenvalue ladder. The uniform-sphere neutron reproduces the
published regions to within $0.25$ decades at $\mu=0.1$ and
$1\,\mathrm{eV}$. At $1\,\mathrm{meV}$ the published boundary lies
$0.1$--$0.8$ decades above ours, while at $1\,\mathrm{keV}$, the upper $\lambda$ boundary differs by up to a factor ${\sim}20$.

Above the mirror, there is an attractive well due to the neutron's
coupling to the square of the symmetron field. The width of this well is set by the symmetron's Compton wavelength in the chamber, which at
$\mu=1\,\mathrm{keV}$ is $m_{\rm bg}^{-1}\simeq0.14\,\mathrm{nm}$,
four orders of magnitude below the characteristic scale
$z_0\simeq5.9\,\mu\mathrm{m}$ of the gravitational states and far
below the resolution of any mesh scaled to them. At a representative point
inside the published exclusion region, $M_s=10^{-3}\,\mathrm{GeV}$ and $\lambda=1$, the resolved
  well has depth scale $V_0=15.9\,\mathrm{meV}$ and surface
  depth $|V_s(0)|=15.1\,\mathrm{meV}$. It binds three states
  of its own, at $-7.99$, $-2.84$, and $-0.370\,\mathrm{meV}$,
  each localized within a few \aa ngstroms of the surface. The gravitational quartet, by contrast, sits at
$1.41$, $2.46$, $3.32$, and $4.08\,\mathrm{peV}$, so the surface states are bound eight to ten orders of magnitude more deeply than the gravitational ladder. The eigenvalue ladder labels them ``levels 1--3'', so the predicted
$E_{13}$ at the point above is a $7.62\,\mathrm{meV}$ gap between two
surface states, far outside the measured peV-scale window, and the point is excluded. 

Surface-localized states also occur at lower masses, as illustrated
by the $\mu=0.1$ and $1\,\mathrm{eV}$ examples in
Appendix~\ref{app:qbounce-states}. At smaller $\mu$, the broader
scalar potential instead modifies the gravitational ladder.
At $1\,\mathrm{keV}$, the surface states are confined to
\aa ngstrom scales and are well separated from the gravitational
states, requiring the resolved-layer treatment described there.

The previously published boundary lies up to a factor ${\sim}20$ in
$\lambda$ above ours.  At a point in the disputed region,
$M_s=10^{-5}\,\mathrm{GeV}$ and $\lambda=10^{2}$, the well is
$0.16\,\mathrm{meV}$ deep, roughly a third of the strength needed to
bind, and holds no states at all. The binding threshold is obtained from the zero-energy
Schr\"odinger equation for the kink potential with a hard wall
at the mirror. In terms of $S=2m_nV_0/m_{\rm bg}^2$, the threshold
is $S_*(w_s)$, approaching $1/2$ when the surface field vanishes.
Here $S\simeq0.15$, well below the threshold.

Its effect on the gravitational spectrum is described by the
scattering length $a_{\rm s}=-0.19\,\mathrm{nm}$. We obtain $a_{\rm s}$ by solving
the zero-energy Schr\"odinger equation in the surface potential,
with $\psi(0)=0$ and gravity neglected. Beyond the well,
$\psi\propto z-a_{\rm s}$, which defines $a_{\rm s}$. To leading order, this is
equivalent to displacing the mirror surface by that distance,
which shifts every level by the same
$2\times10^{-5}\,\mathrm{peV}$ and cancels in the transition
energies, so the predicted transitions stay at their unperturbed
values, $E_{13}=1.915$ and $E_{14}=2.676\,\mathrm{peV}$, inside both
measured windows, under every assignment of levels to the
eigenvalues.  The level convention therefore cannot account for the difference from the published analysis in this region, which must arise in the computed spectra. An unresolved \aa ngstrom-scale layer is a natural
candidate, but we have not reproduced the published calculation, and
which implementation detail is responsible remains open.

These results rest on a critical assumption. The level
convention here reads the measured $E_{13}$ (and $E_{14}$) as a transition out of
whichever state is lowest, so that at $1\,\mathrm{keV}$ an
\aa ngstrom-scale state, orders of magnitude more deeply bound than
the states the experiment prepares and drives, inherits the measured
transitions.  Whether such a state actually participates in the measurement
at all, whether the state selector populates it, the mechanical
drive addresses it, or the detector counts it, is worthy of further investigation.

The source shape becomes important toward stronger coupling, including in the
lower-$\mu$ panels. At $\mu=1\,\mathrm{keV}$ and
$M_s=\ConstraintSymQbounceMassKev\,\mathrm{keV}$, the strongest
coupling at which all three sources remain inside the display cuts,
the sphere, chiral and pion-truncated upper $\lambda$ limits are
$\ConstraintSymQbounceSphereLambda$, $\ConstraintSymQbounceChiralLambda$
and $\ConstraintSymQbouncePionLambda$, gains of
$\ConstraintSymQbounceChiralGain$ and $\ConstraintSymQbouncePionGain$
at fixed $M_s$, falling to below a factor $10$ by $M_s\sim1\,\mathrm{MeV}$.

\paragraph{Neutron interferometry.}
We evaluate both interferometry modes in the shared finite-cylinder model 
(Sec.~\ref{sec:observables-interferometry}), and for the
uniform-sphere neutron the published boundaries lie above the
recomputed ones by a median of $0.04$ decades at $1\,\mathrm{keV}$,
$0.09$ decades at $1\,\mathrm{eV}$ and $0.14$ decades at $0.1\,\mathrm{eV}$,
rising toward the symmetry-restoration edge. At first order the phase observable is linear in
$\Q/\lambda$, and the displayed regions terminate, like the
published ones, at the strong-coupling edge $A-1 = 0.02$.

This termination is independent of the small-potential condition
behind Eq.~\eqref{eq:interferometry-phase}. At the $A-1$ edge the
potential envelope $V_{\rm env}=\Q_n m_n\phibg^2/(2M_s^2)$, with
$\phibg^2=(\mu^2-\rbg/M_s^2)/\lambda$ the broken-phase value
in the evacuated cell, reaches $0.02\,\Q_n m_n$, which exceeds $E_{\rm kin}$ wherever
$\Q_n\gtrsim6\times10^{-10}$. We therefore distinguish three regimes
in Fig.~\ref{fig:constraints-symmetron}. Only
$V_{\rm env}\le0.1E_{\rm kin}$ is shown as a linear-phase exclusion.
For $0.1E_{\rm kin}<V_{\rm env}<E_{\rm kin}$ the beam still propagates
but the phase must be evaluated beyond first order, and for
$V_{\rm env}\ge E_{\rm kin}$ transmission itself is unverified.

Ref.~\cite{Fischer:2023eww} uses a pressure-mode
threshold of $3.5^\circ$, fillings of $10^{-2}$ and
$2.4\times10^{-4}\,\mathrm{mbar}$, and the field set to its vacuum
value $\phi_{\rm v}$ throughout the cell. We use the one-sided $95\%$ limit
$5.44^\circ$ from the measured $\gamma=0.37^\circ\pm3.53^\circ$ of
Ref.~\cite{Lemmel:2015kwa}, which places our boundaries $0.19$ decades
lower at every $\mu$. The flat-field approximation is accurate to
better than $1\%$ along these boundaries in our finite-cylinder
solution, away from the symmetry-restoration edge. At $\mu=1\,\mathrm{meV}$ the published boundary lies a
further factor of up to $5$ above ours near $M_s\sim1\,\mathrm{GeV}$,
falling to about $1.4$ by $M_s\sim10^{3}\,\mathrm{GeV}$. For $M_s\gtrsim1\,\mathrm{GeV}$
the neutron is unscreened, both fillings are in the broken phase, and
the pressure-mode signal reduces to $\Q_n m_n\Delta\rho/(2\lambda M_s^4)$,
independent of $\mu$, so our boundaries at the three slices coincide
there. The published $1\,\mathrm{meV}$ contour instead sits a factor
$5$ above the published $0.1\,\mathrm{eV}$ contour at $M_s=1\,\mathrm{GeV}$.

The increase in the area of the excluded region for a Skyrmion model compared to a uniform sphere is large for neutron interferometry. The exclusion boundary lies well inside the screened regime, and at its apex the
neutron's screening parameter is $s\simeq3\times10^{3}$ at
$\mu=1\,\mathrm{eV}$ and $3\times10^{6}$ at $1\,\mathrm{keV}$. The
maximal reach in $\lambda$ rises by $\ConstraintSymInterferometryChiralEv$ decades at
$\mu=1\,\mathrm{eV}$ and by $\ConstraintSymInterferometryChiralKev$ decades at $1\,\mathrm{keV}$.  The
physical pion mass suppresses this somewhat, to $\ConstraintSymInterferometryPionEv$ and $\ConstraintSymInterferometryPionKev$ decades respectively, because the field here penetrates much farther beyond the pion cutoff than at the chameleon qBOUNCE edge. The turning radius $a\sqrt s$ is $\simeq60$ and $1.8\times10^{3}$ core radii, compared with a cutoff at $2.3\,a$, well inside the regime where the
enhancement grows only logarithmically. All these maxima lie well inside the linear-phase domain, with $V_{\rm env}/E_{\rm kin}\sim10^{-10}$.

\paragraph{Hydrogen and muonium.}
For the uniform sphere proton the recomputed hydrogen and muonium
regions track the published contours. One input differs by
construction. The published analysis screens the proton with the
interpolation $\min(3/\kappa^2,1)$, while we evaluate the full
Brax--Pitschmann factor. The two coincide in the unscreened and
deeply screened limits and differ only where the proton is marginally
screened, by up to a factor $2.2$ in charge near
$\kappa\simeq\sqrt3$, which moves our boundaries by at most half a
decade in $\lambda$ over a few mass columns at the weak-coupling
corner. The Skyrmion moves the strong-coupling edges by the proton's charge
ratio, since the spectroscopy observable is linear in
$\Q_p/\lambda$. At $\mu=1\,\mathrm{keV}$ the hydrogen reach in
$\lambda$ rises by $\ConstraintSymHydrogenPionKev$ decades with the physical tail and by $\ConstraintSymHydrogenChiralKev$
in the chiral limit.  The muonium regions are again the same for all three source models.

\section{Discussion}
\label{sec:discussion}

Scalar fields are popular as models of dynamical dark energy  and ultra-light dark matter. If such fields are introduced, an explanation is needed for why they have not yet been detected.  Screening mechanisms provide an alternative to stringent fine-tuning of coupling constants, by allowing the properties of scalar fluctuations to vary with their environment. This comes at the cost of making the theory non-linear. 

Whilst screening mechanisms were introduced to avoid local experimental tests, they do not render the scalar invisible. Precision measurements with small and compact objects have proved a powerful probe of these theories. As the sensitivity of these experiments continues to improve, it is vital that the theoretical accuracy of our predictions for the sensitivity of the experiments is refined as well. It is in this spirit that in this work we revisit the assumptions that have been made in existing analyses about the way in which the scalar field behaves around neutrons and protons.

Protons and neutrons are complex objects made of quarks and gluons. When probed at low energies much of this complexity can be neglected, but when computing the effect of such nucleons on screened scalar fields the spatial distribution of mass and energy is important.  In this work we have shown that different models of the nucleon make a significant difference to the regions of chameleon and symmetron parameter space excluded by such experiments. Our two models were a hard sphere of constant density and fixed radius, and a Skyrmion profile, which we believe to be a more accurate representation. All preceding analyses assumed that the nucleon was a hard sphere. 

Figures \ref{fig:constraints-chameleon} and \ref{fig:constraints-symmetron} show the consequences for constraints on the chameleon and symmetron models respectively, identifying the parts of parameter space in which the constrained region shifts when the model of the nucleon is changed. Key changes include the fact that  symmetron constraints from neutron interferometry at $\mu=1\,\mathrm{keV}$ reach
maximal self-couplings about $\ConstraintSymInterferometryPionKev$ orders of magnitude higher than for a
uniform-sphere neutron ($\ConstraintSymInterferometryChiralKev$ orders of magnitude in the chiral limit),
and at the dark-energy scale the hydrogen $1s$--$2s$ chameleon bound
reaches couplings $\ConstraintHydrogenDEGainPion$ orders of magnitude
weaker than for a uniform-sphere proton ($\ConstraintHydrogenDEGainChiral$
in the chiral limit), overtaking the muonium bound.
We propose that, in future, the Skyrmion model be used to analyse constraints on chameleons and symmetrons, and related screening models, from precision measurements of protons and neutrons. 

We have not addressed here the interesting question of how best to model the nuclei of more complex atoms, such as those used in atom interferometry experiments which also provide important constraints on the type of screening models discussed here.  Similar improvements in the accuracy of theoretical predictions are needed to continue to ensure that forecasts match the precision of experiments. Determining how best to model the screening of such atomic nuclei is left as a question for future work. 

\section*{Acknowledgments}

CB and AM are supported by  STFC Consolidated Grants  [Grant Nos.  ST/X000672/1, ST/C004235/1]. For the purpose of open access, the authors have applied a CC BY public copyright license to any Author Accepted Manuscript version arising. We acknowledge the use of Anthropic's Claude Fable in this work, for the derivation of the near-zone symmetron charge, Eq.~\eqref{eq:trace-near-zone-charge}. This result was subsequently checked by a human researcher.

\section*{Data availability}

The screening-charge tables, the recomputed exclusion boundaries and the figure files supporting this work are openly available on Zenodo at \url{https://doi.org/10.5281/zenodo.23419321} under a CC BY 4.0 license. The analysis code is available from the authors on reasonable request.

\appendix

\section{Derivation of the Symmetron profile}
\label{app:symmetron}

\emph{Uniform sphere.}---For the uniform sphere, where the density in the interior of the sphere restores the symmetry of the symmetron model, 
the solution of the linearized field equation regular at the origin is
$\phi_{\rm in}=A\sinh(k_+r)/r$ with $k_+^2=\rhos/M_s^2-\mu^2$, and the
exterior linearized solution is
$\phi_{\rm out}=\phibg-(D/r)\,e^{-m_{\rm bg}(r-\rs)}$ with
$m_{\rm bg}^2=2\left(\mu^2-\rbg/M_s^2\right)$. Writing $x=k_+\rs$ and
$y=m_{\rm bg}\rs$, continuity of the field and its derivative at $\rs$
determines the exterior coefficient,
$D=\phibg\rs\,(1-\tanh x/x)/(1+y\tanh x/x)$, with the ratio of the two
conditions eliminating the interior normalization. Normalizing $D$ by
its leading small-$x$ expression at fixed $y$, the unscreened limit
$D_0=\phibg\rs x^2/[3(1+y)]$, gives the Brax--Pitschmann screening
factor~\cite{Brax:2017hna,Pitschmann:2020ejb}:
\begin{equation}
  \Q_{\rm BP}=(1+y)\,\frac{3}{x^2}\,
 \frac{1-\tanh x/x}{1+y\tanh x/x},
 \label{eq:bp-restored}
\end{equation}
whose massless-near-zone limit $y\to0$, $x\to\kappa$ is
Eq.~\eqref{eq:sphere-near-zone-charge}.  The point-source reference
$D_{\rm point}=\phibg\Mcal/(4\pi M_s^2)=\phibg\rs\kappa^2/3$ instead
gives
\begin{equation}
 \frac{D}{D_{\rm point}}
 =\Q_{\rm BP}\frac{x^2}{\kappa^2(1+y)},
 \qquad D_Y=e^yD,
 \label{eq:sphere-monopole-conversion}
\end{equation}
where $D_Y$ multiplies $e^{-m_{\rm bg}r}/r$. The origin-referenced,
point-normalized coefficient is therefore $D_Y/D_{\rm point}$, not
$\Q_{\rm BP}$. The two coincide only in the massless near zone.

When the sphere density is too
low to restore the symmetry, $\rhos<\mu^2M_s^2$, the interior ansatz
$\phi_{\rm in}=\phi_{\rm m}+A\sinh(k_-r)/r$ with $k_-^2=2(\mu^2-\rhos/M_s^2)$ and $\phi_{\rm m}^2=(\mu^2-\rhos/M_s^2)/\lambda$ gives in the same way~\cite{Pitschmann:2020ejb}
\begin{equation}
  \Q_{\rm BP}=\frac{3\,(1+y)\left(x-\tanh x\right)}
 {x^2\left(x+y\tanh x\right)},
 \qquad x=k_-\rs.
 \label{eq:bp-broken}
\end{equation}
In this branch $D$ and $D_0$ both carry $\phibg-\phi_{\rm m}$ in place
of $\phibg$. The offset cancels in $\Q_{\rm BP}$ but leaves the
point-normalized coefficient of
Eq.~\eqref{eq:sphere-monopole-conversion} with a further factor
$1-\phi_{\rm m}/\phibg$.

\emph{Trace source.}---In the massless near zone the trace source obeys
Eq.~\eqref{eq:symmetron-near-zone} with
$\rho=\rho_A\min\!\left[1,(a/r)^6\right]$. Regularity gives
$r\phi\propto\sinh(sr/a)$ in the core.  In the tail, the inversion
$z=a/r$ with $\chi(z)=\phi(a/z)$ reduces the radial equation for $r\phi$
to $\chi''=s^2z^2\chi$, solved by
$\sqrt z\,I_{\pm1/4}(sz^2/2)$.  As $z\to0$,
\begin{align}
 \sqrt z\,I_{-1/4}\!\left(\tfrac{sz^2}{2}\right)&\to
 \frac{(s/4)^{-1/4}}{\Gamma(3/4)},
 \nonumber\\
 \sqrt z\,I_{+1/4}\!\left(\tfrac{sz^2}{2}\right)&\to
 \frac{(s/4)^{1/4}}{\Gamma(5/4)}\,z,
 \label{eq:bessel-normalizations}
\end{align} 
so the $I_{-1/4}$ branch carries the asymptotic field and the $I_{+1/4}$
branch carries the exterior monopole coefficient $D\propto z=a/r$.  The
derivatives at the core edge collapse via the Bessel recurrences to
\begin{align}
 \frac{\dd}{\dd z}\left[\sqrt z\,
 I_{-1/4}\!\left(\tfrac{sz^2}{2}\right)\right]_{z=1}
 &=s\,I_{3/4}\!\left(\tfrac{s}{2}\right),\\
 \frac{\dd}{\dd z}\left[\sqrt z\,
 I_{+1/4}\!\left(\tfrac{sz^2}{2}\right)\right]_{z=1}
 &=I_{1/4}\!\left(\tfrac{s}{2}\right)+s\,I_{5/4}\!\left(\tfrac{s}{2}\right).
\end{align}

Matching $\phi$ and $\phi'$ at $r=a$ and normalizing by
$D_{\rm point}=\phibg\Mcal/(4\pi M_s^2)=\tfrac23\,s^2a\,\phibg$ gives
Eq.~\eqref{eq:trace-near-zone-charge}.  The $\Gamma$-function prefactor is
the ratio of the two normalizations in
Eq.~\eqref{eq:bessel-normalizations}, written using
$\Gamma(1/4)=4\,\Gamma(5/4)$, and the unscreened limit $\Q\to1$ follows
identically from them.

\section{The screening charge in a non-empty background }
\label{app:symmetron-flux}

Beyond the near zone the exterior field is Yukawa suppressed,
and the  coefficient of the suppression depends on the matching convention. To extend the
near-zone characterization to finite background mass, we must therefore analyse how the screening can be defined in a background where the symmetron becomes massive. This should also apply to cases where the density is not that of a sphere but slowly decreases away from the objects. 

In a non-empty background, the field acquires both a mass $m_{\rm bg}$
and a value far away from the object $\phibg$. In a near-zone interval
outside a localized source, with $m_{\rm bg}r\ll1$, the field profile
simplifies to
\begin{equation}
 \phi\sim\phibg-\frac{D}{r},
\end{equation}
with the corresponding point-source description
\begin{equation}
 \nabla^2\phi=4\pi D\,\delta^{(3)}(\bm r),
\end{equation}
so that the field seems to emanate from a point with a source
determined by $D$. For a true point particle in this background,
\begin{equation}
 4\pi D_{\rm point}=\Mcal\partial_\phi A(\phibg),
\end{equation}
and the near-zone screening charge is $\Q=D/D_{\rm point}$.

Beyond the near zone the exterior is no longer harmonic, and the
monopole coefficient depends on how the Yukawa factor is referenced
(Appendix~\ref{app:symmetron}). We therefore replace it by the flux
through a sphere of radius $R$,
\begin{equation}
 L(R)\equiv R^2\phi'(R)
 =\int_{R_*}^R\partial_\phi V_{\rm eff}(\phi(r);r)\,r^2\,\dd r,
\end{equation}
where the second equality integrates the spherical equation of motion
\begin{equation}
 \frac{\dd}{\dd r}\left(r^2\frac{\dd\phi}{\dd r}\right)
 =r^2\partial_\phi V_{\rm eff}(\phi(r);r)
 \label{eq:symmetron-radial-eom}
\end{equation}
from the radius $R_*$ at which $\phi'(R_*)=0$. In the massless near
zone $\phi'=D/r^2$, so $L(R)=D$ is independent of $R$, while at finite
background mass the radius dependence is explicit. The screening
charge is then generalized to
\begin{equation}
 \Q_F(R;\phibg)\equiv\frac{L(R)}{D_{\rm point}}
 =\frac{4\pi\int_{R_*}^R
 \partial_\phi V_{\rm eff}(\phi(r);r)\,r^2\,\dd r}
 {\partial_\phi A(\phibg)\Mcal},
 \label{eq:symmetron-flux}
\end{equation}
which reduces to $\Q=D/D_{\rm point}$ wherever the flux is
approximately constant. The normalization uses the fixed total source
mass $\Mcal$, not the mass enclosed within $R$.

For a thin-shell solution, $R_*$ is the screening radius below which
$\phi\simeq\phi_c$, the minimum of the interior effective potential.
Writing
\begin{equation}
 J(r)\equiv\int_{R_*}^r
 \partial_\phi V_{\rm eff}(\phi(u);u)u^2\,\dd u,
\end{equation}
integration of the field equation gives the condition
\begin{equation}
 \phibg-\phi(R_*)=\int_{R_*}^\infty\frac{J(r)}{r^2}\,\dd r.
\end{equation}
For chameleons this is the screening-radius construction with
$\phi(R_*)\simeq\phi_c$. Neglecting the potential contribution in
the thin shell and outside the body, Eq.~\eqref{eq:symmetron-flux}
reduces to Eq.~\eqref{eq:general-thin-shell}, with
$\partial_\phi A\simeq M_c^{-1}$. For a regular symmetron solution we
instead take $R_*=0$, since regularity gives $\phi'(0)=0$, without requiring
the central field to equal the interior minimum exactly.
In the massless near zone the normalized symmetron charge is independent
of $\phibg$. This allows the neutron energy in the mirror's field to be
written as $\Q^{\rm nz}m_n\phi^2/(2M_s^2)$, up to a reference constant.
At finite radius the normalized equation depends on the environment
only through $m_{\rm bg}$ and the density contrast
$\delta\rho=\rho_{\rm tot}-\rbg$. No independent field normalization enters.

In the general case, Eq.~\eqref{eq:symmetron-flux} has the disadvantage
that $R_*$ needs to be known. For a massive homogeneous exterior,
$r^2\phi'(r)\to0$ at infinity, so the field equation also gives
\begin{equation}
 \Q_F(R;\phibg)=-\frac{4\pi\int_R^\infty
 \partial_\phi V_{\rm eff}(\phi(r);r)r^2\,\dd r}
 {\partial_\phi A(\phibg)\Mcal},
\end{equation}
which only depends on the field profile outside the evaluation radius.

For the matched sphere profiles of Appendix~\ref{app:symmetron},
$L(\rs)=(1+y)D$, hence
\begin{align}
 \Q_F(\rs)&=\frac{(1+y)D}{D_{\rm point}}\nonumber\\
 &=\frac{x^2}{\kappa^2}
 \begin{cases}
  \Q_{\rm BP},&\text{restored interior},\\
  (1-\eta)\Q_{\rm BP},&\text{broken interior}.
 \end{cases}
\end{align}
Here $\eta=\phi_{\rm m}/\phibg$. In the restored branch, dropping
the interior $-\mu^2$ term sets $x=\kappa$ and gives
$\Q_F(\rs)=\Q_{\rm BP}$. 

\paragraph{Radial flux of the trace source.}

\begin{figure}[tbp]
 \centering
 \includegraphics[width=\columnwidth]{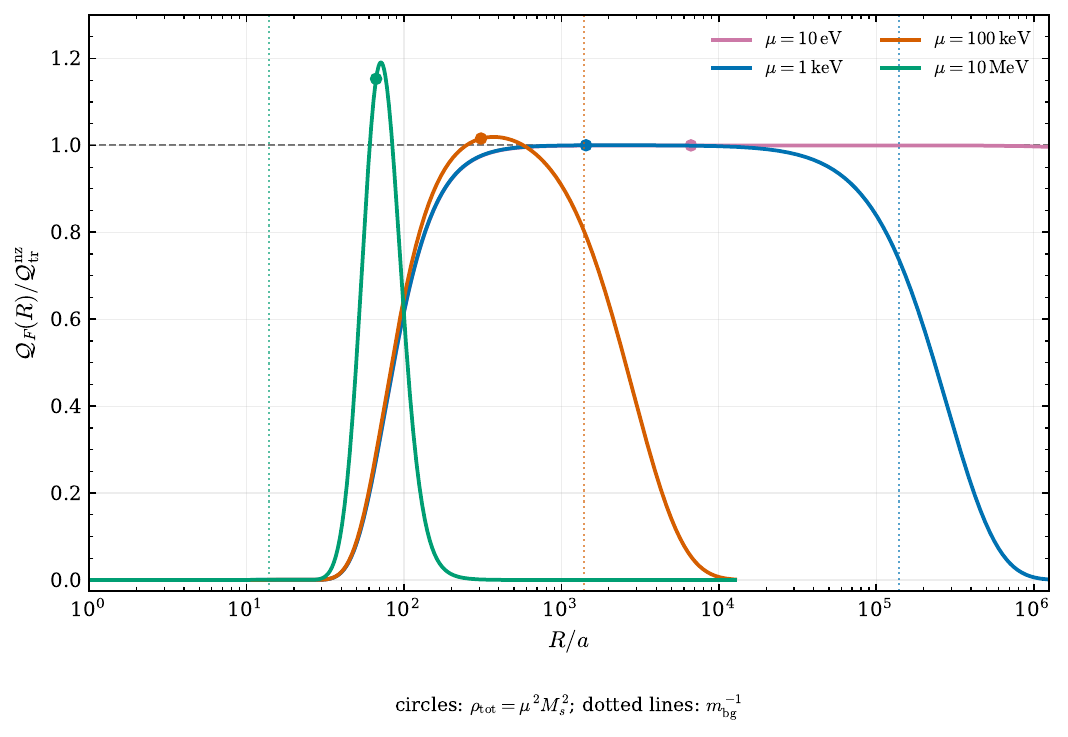}
 \caption{Radial dependence of the symmetron flux, Eq.~\eqref{eq:symmetron-flux},
 evaluated on full nonlinear trace solutions and normalized to the analytical
 near-zone charge. All curves use $a=\BenchmarkCoreRadiusFm\,\mathrm{fm}$,
 $\Mcal=\BenchmarkCoreMassMeV\,\mathrm{MeV}$, $M_s=10\,\mathrm{keV}$ and
 $\rbg=\BenchmarkBackgroundDensity\,\mathrm{kg/m^3}$, differing only in $\mu$.
 The dashed horizontal line denotes agreement with the near-zone result;
 circles mark $\rho_{\rm tot}(R_\rho)=\mu^2M_s^2$ and dotted vertical lines
 mark $m_{\rm bg}^{-1}$. The $10\,\mathrm{eV}$ Compton scale lies beyond
 the plotted range. The $100\,\mathrm{keV}$ and $10\,\mathrm{MeV}$
 curves are displayed only to $1.3R_{\rm floor}$.}
 \label{fig:symmetron-flux}
\end{figure}

Figure~\ref{fig:symmetron-flux} tests this interpretation on full nonlinear
trace solutions, holding the source and background density fixed. At
$\mu=10\,\mathrm{eV}$ and $1\,\mathrm{keV}$, broad intervals agree
with the near-zone charge, extending farther in radius for the lighter case.
The interval narrows at $100\,\mathrm{keV}$, and at $10\,\mathrm{MeV}$ there is no near-zone plateau, only a pronounced maximum.

A natural candidate evaluation radius is the density-crossing radius
$R_\rho$, where $\rho_{\rm tot}(R_\rho)=\mu^2M_s^2$ and the quadratic
term in the effective potential changes sign. For our floored trace tail,
\begin{equation}
 R_\rho=a\left(\frac{\rho_A}{\mu^2M_s^2}\right)^{1/6},
 \qquad \rbg<\mu^2M_s^2<\rho_A.
\end{equation}
Outside this interval there is no unique tail crossing. At a crossing,
$V_{{\rm eff},\phi}=\lambda\phi^3$, so
\begin{equation}
 \left.\frac{\dd\Q_F}{\dd R}\right|_{R_\rho}
 =\frac{4\pi M_s^2\lambda}{\Mcal\phibg}R_\rho^2\phi(R_\rho)^3>0
\end{equation}
on the positive nontrivial branch. The crossing does not exactly terminate
scalar sourcing. Its marker lies on the plateaus in the two lightest benchmarks,
but not on a broad plateau in the heavier cases.

Beyond $R_\rho$, the source's density tail can still affect the field.
Once the density reaches the background floor at
$R_{\rm floor}=a(\rho_A/\rbg)^{1/6}$ and the field is sufficiently close
to $\phibg$, the usual Yukawa form applies:
$\phi-\phibg=-C e^{-m_{\rm bg}R}/R$. This gives
\begin{equation}
 \Q_F(R)=\frac{4\pi M_s^2 C}{\Mcal\phibg}
 (1+m_{\rm bg}R)e^{-m_{\rm bg}R}.
\end{equation}
The resulting decline in flux is a finite-range effect and does not represent additional screening of the source.

\paragraph{Near-zone validation.}

Full nonlinear source solutions support the near-zone approximation for
the neutron and proton sizes used in the constraints. We sample broken
backgrounds at the laboratory gas densities and in the negligible-density
limit, with $10^{-4}\leq\mu/\mathrm{eV}\leq10^3$ and
$1\leq M_s/\mathrm{eV}\leq10^{24}$.
For the chiral trace, $\Q_F(R)$ agrees with $\Q_{\rm tr}^{\rm nz}$
to within $\ConstraintFluxTracePercent\%$ throughout $R_0\leq R\leq2R_0$, where
$R_0=a(50/\Q_{\rm tr}^{\rm nz})^{1/3}$ is chosen so the remaining
matter contribution is bounded by $1\%$ of the near-zone charge.
The largest departure occurs for the proton-sized source at
$\mu=1\,\mathrm{keV}$ and $M_s=1\,\mathrm{eV}$ in the
negligible-density background. For the uniform sphere, the nonlinear surface flux, BP prescription and
$\Q_{\rm sph}^{\rm nz}$ agree within $\ConstraintFluxSphereBound$ fractionally
across the same scan, including both interior branches.

\section{qBOUNCE spectra at the mirror}
\label{app:qbounce-states}

  This appendix examines the qBOUNCE eigenvalue spectra used to obtain the symmetron constraints in Sec.~\ref{sec:constraints-symmetron}.
  We present representative energy levels and transition energies in the broad, intermediate, and thin-layer regimes, and test the numerical convergence of the methods used in each regime.  Throughout, levels are numbered in order of increasing energy, following the convention used in the constraint analysis.

  The stationary Schr\"odinger equation for a neutron above a single mirror is
  \begin{equation}
   \left[-\frac{1}{2m_n}\frac{\dd^2}{\dd z^2}
   +m_ngz+V_s(z)\right]\psi
   =E\psi,
   \qquad \psi(0)=0\,.
   \label{eq:qbounce-hamiltonian}
  \end{equation}
The boundary condition $\psi(0)=0$ treats the mirror as a hard wall,
following the published
analyses~\cite{Cronenberg:2018qxf,Brax:2017hna}. The glass mirror in fact
presents a finite optical potential, into which ultracold neutrons
penetrate to a depth $d\sim10\,\mathrm{nm}$. For the gravitational states
alone this changes the transition frequencies by only a few parts in
$10^{9}$~\cite{Cronenberg:2018qxf}. In our slices with
$\mu\le1\,\mathrm{eV}$, the scalar layer is also much wider than $d$.
This separation of scales is lost when the layer width
$\ell=m_{\rm bg}^{-1}$ falls below $d$, at
$\mu\gtrsim14\,\mathrm{eV}$ for $m_{\rm bg}\simeq\sqrt{2}\mu$.
The hard-wall approximation is therefore not well justified by this
scale separation at $\mu=1\,\mathrm{keV}$. We retain it for
consistency with the published treatment and return to this
caveat at the end of the appendix.
  
  For the one-mirror symmetron kink, the localized surface potential is
  \begin{equation}
   V_s(z)=-V_0\,\frac{4w}{(1+w)^2},
   \qquad w=w_s\,e^{-m_{\rm bg}z},
   \label{eq:layer-well}
  \end{equation}
  where $V_0=\Q\,m_n(A-1)$ sets the depth scale and is evaluated at the residual-gas density, while $w_s=w(0)$ is the surface value of $w$.
  The spatially constant asymptotic contribution to $\delta V$ has been subtracted, since it shifts all energy levels equally and therefore cancels from transition energies.

  Gravity defines the characteristic length $z_0=(2m_n^2g)^{-1/3}=5.87\,\mu\mathrm{m}$ and energy $E_0=m_ngz_0=0.602\,\mathrm{peV}$.  In the absence of the scalar potential, the four lowest levels are $1.41$, $2.46$, $3.32$, and $4.08\,\mathrm{peV}$.  The scalar well is instead characterized by its width $\ell=m_{\rm bg}^{-1}$ and depth scale $V_0$. Here $m_{\rm bg}$ is the broken-phase symmetron mass at the residual-gas density above the mirror, $m_{\rm bg}^2=2\left(\mu^2-\rbg/M_s^2\right)$ as in Sec.~\ref{sec:symmetron}, which reduces to $m_{\rm v}=\sqrt{2}\,\mu$ in the negligible-density limit.

  The relative sizes of $\ell$ and $z_0$ guide the numerical
     treatment, with the ratio $\ell/z_0$ separating two numerical regimes.  When the
  surface layer is resolved on the gravitational-scale mesh, we solve
  Eq.~\eqref{eq:qbounce-hamiltonian} directly. At the two $\mu=1\,\mathrm{keV}$ worked points considered below,
  $\ell=0.14\,\mathrm{nm}$ and $\ell/z_0\simeq2.4\times10^{-5}$.
  The gravitational potential energy changes by only
  $m_ng\ell=1.4\times10^{-5}\,\mathrm{peV}$ across the layer, so the
  near-surface and gravitational dynamics are well separated.  This
  motivates a two-zone treatment: we solve the surface layer in its own
  natural units, retain any bound states that it supports, and describe
  its effect on the outer gravitational spectrum through its scattering
  length.

  Introducing the dimensionless coordinate $x=z/\ell$ and eigenvalue
  $\varepsilon=E/E_u$, where
  \begin{equation}
   E_u=\frac{m_{\rm bg}^2}{2m_n},
  \end{equation}
  the gravitational term in the rescaled Schr\"odinger equation is
  $(\ell/z_0)^3x$.  At $\mu=1\,\mathrm{keV}$ its coefficient is
  approximately $1.3\times10^{-14}$, and is therefore negligible over
  the surface layer, where $x=\order(1)$.  Dropping this term gives the near-surface (inner) problem 
  \begin{equation}
   -\psi''-S\,\frac{4w}{(1+w)^2}\psi
   =\varepsilon\psi,
   \qquad
   w=w_s e^{-x},
   \qquad
   \psi(0)=0,
   \label{eq:layer-problem}
  \end{equation}
  where primes denote derivatives with respect to $x$ and 
   \begin{equation}
   S=\frac{2m_nV_0}{m_{\rm bg}^2}.
  \end{equation}
   We solve Eq.~\eqref{eq:layer-problem} numerically on the half-line using
  the full kink potential.  Negative eigenvalues $\varepsilon_n$
  correspond to surface-bound states, with physical energies
  \begin{equation}
   E_n^{\rm layer}=E_u\varepsilon_n.
  \end{equation}
  In the coordinate $x$, the layer has width of order unity and can
  therefore be resolved without simultaneously discretizing the
  micrometre-scale gravitational region.

  For the gravitational states,
  $\varepsilon=E/E_u=\order[(\ell/z_0)^2]\ll1$, so their response to the
  surface layer is determined by the zero-energy inner solution
  $\psi_0$.  Once the kink potential has decayed, this solution takes
  the form $\psi_0\propto z-a_{\rm s}$, which defines the scattering length
  $a_{\rm s}$.  At a matching point in the overlap region
  $\ell\ll z\ll z_0$, the surface potential is negligible while gravity
  has not yet modified this linear behaviour.  We therefore equate the
  logarithmic derivatives of the inner and outer solutions there. At the two $1\,\mathrm{keV}$ worked points, $|a_{\rm s}|\ll z_0$,
     and the leading effect of this matching is equivalent to
     displacing the outer hard wall to $z=a_{\rm s}$.  The resulting
     gravitational spectrum is the Airy ladder shifted by the
     common amount $m_nga_{\rm s}$.  This shift has magnitude
     $\lesssim10^{-4}\,\mathrm{peV}$ at these points and cancels
     from gravitational transition energies at this order.   The reconstructed two-zone spectrum consists of the surface-bound
  levels together with this shifted gravitational ladder.  This
  separation requires the surface-bound states to remain localized
  well within the gravitational length scale. States close to a
  binding threshold require separate examination.

    The spectra presented below are evaluated at exact worked points using
  the same methods as the constraint grids: direct diagonalization of
  Eq.~\eqref{eq:qbounce-hamiltonian} for $\mu\leq1\,\mathrm{eV}$ and
  the two-zone reduction for $\mu=1\,\mathrm{keV}$.  The direct
  calculation uses a graded mesh of $700$ nodes extending to
  $180\,\mu\mathrm{m}$, with cell widths increasing from
  $0.26\,\mathrm{nm}$ at the mirror to $0.8\,\mu\mathrm{m}$ at the
  top of the domain.  The grid is finest near the mirror to resolve the narrow scalar well,
and becomes coarser farther away to cover the gravitational states.
Besides requiring at least four interior mesh nodes within one interaction
range, we check that the local mesh spacing resolves the wavefunction's
oscillations by monitoring $\max_z[k(z)\Delta z(z)]$, where $k(z)$ is
the local wave number and $\Delta z(z)$ is the local mesh spacing.

The mesh must resolve not only the width of the well but also
the neutron's oscillations within it. A useful measure is the
zero-energy de~Broglie wavelength at the bottom of the well,
  \begin{equation}
   \lambda_{\rm dB}
   =\frac{2\pi}{\sqrt{2m_n|V_s(0)|}}.
  \end{equation}
   At the excluded worked points with
  $M_s=10^{-3}\,\mathrm{GeV}$ and $\lambda=1$,
  $\lambda_{\rm dB}$ is
  $2.4\,\mu\mathrm{m}$ for $\mu=0.1\,\mathrm{eV}$,
  $0.23\,\mu\mathrm{m}$ for $\mu=1\,\mathrm{eV}$, and
  $0.23\,\mathrm{nm}$ for $\mu=1\,\mathrm{keV}$.  The surface layer
  contains $653$, $139$, and $61$ interior mesh nodes for
  $\mu=1\,\mathrm{meV}$, $0.1\,\mathrm{eV}$, and
  $1\,\mathrm{eV}$, respectively. At the $1\,\mathrm{keV}$ worked points, however,
  $\ell\simeq0.14\,\mathrm{nm}$ is smaller than the first
  $0.26\,\mathrm{nm}$ cell, so no interior mesh nodes lie
  within one interaction range of the mirror.

To test the two-zone approximation, we also solve the full
Schr\"odinger equation with both the surface potential and gravity
on a single uniform grid, and compare the surface-bound energies.  For the model
  $M_s=10^{-3}\,\mathrm{GeV}$, $\lambda=1$, the three deeply bound
  levels agree to within a relative difference of $2\times10^{-4}$
  at $\mu=1$, $10$, and $100\,\mathrm{eV}$.
  At $1\,\mathrm{keV}$, a uniform mesh of $6.6\times10^6$ nodes
  over $40\,\mu\mathrm{m}$ gives agreement within $0.50\%$.
  Increasing the node count from $4.4\times10^6$ to
  $6.6\times10^6$ changes the shallowest of these three levels
  by $0.66\%$.  These comparisons test the surface-bound energies. 
  The gravitational ladder is obtained through the outer matching
  described above.

  For the displayed slices with $\mu\leq1\,\mathrm{eV}$, we also
  compare the $700$-node production spectra with refined uniform-grid
  solutions of the full equation.  At the worked points, the five
  lowest levels agree within $0.4\%$, except for the shallow fourth
  state at the excluded $1\,\mathrm{eV}$ point, which differs by
  $5.4\%$.  The allowed comparison points use the same $M_s$ as the
  corresponding excluded rows of Table~\ref{tab:qbounce-levels},
  but with $\lambda=1$, $10^3$, and $10^2$ at
  $\mu=1\,\mathrm{meV}$, $0.1\,\mathrm{eV}$, and
  $1\,\mathrm{eV}$, respectively. At the allowed worked points, the production values of
  $E_{13}$ differ from the reference solutions by less than
  $1.2\times10^{-4}\,\mathrm{peV}$, well below the experimental
  half-width of $1.1\times10^{-2}\,\mathrm{peV}$.  Doubling the
  resolution of the uniform reference grids changes $E_{13}$ at
  these allowed points by less than $10^{-6}\,\mathrm{peV}$.  At the tested points, these numerical differences are too small
to change whether the predicted transitions lie inside or outside
the experimental windows.

  \begin{figure*}[tbp]
 \includegraphics[width=\textwidth]{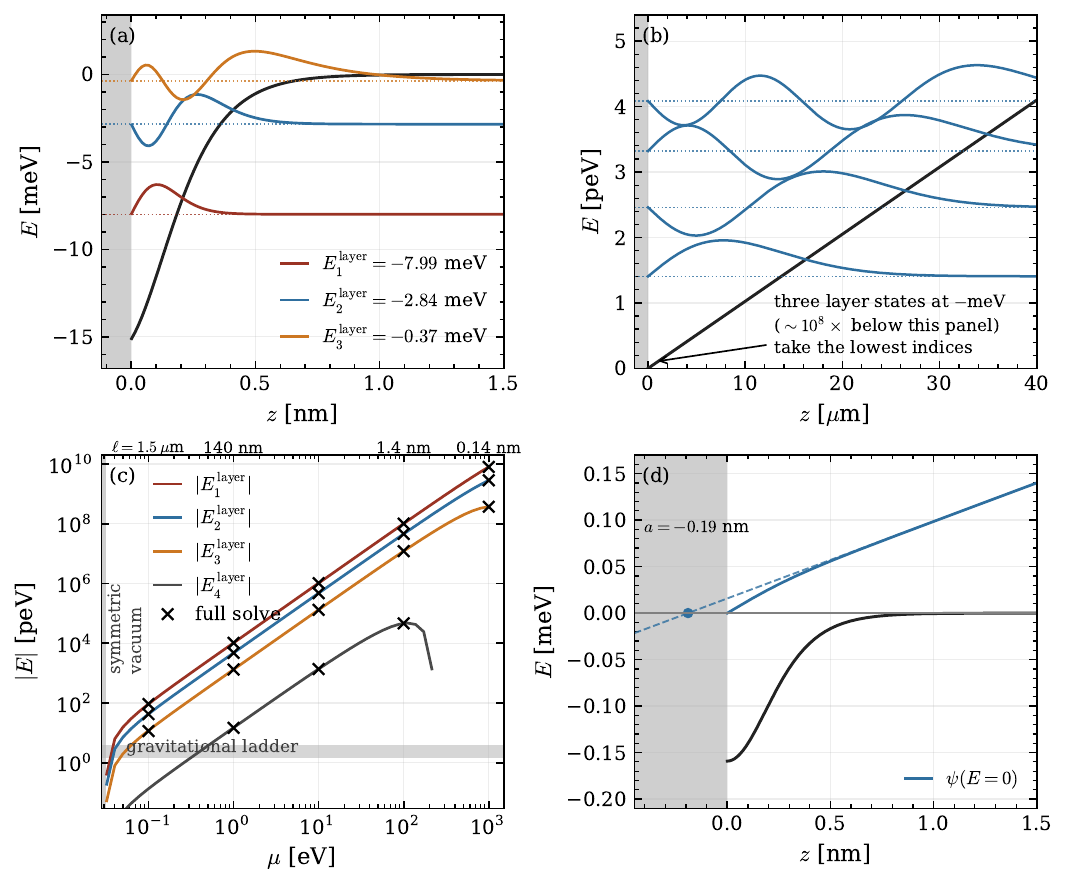}
 \caption{qBOUNCE symmetron spectra at the mirror, for the neutron
 sphere charge.  (a)~The resolved layer at the excluded point
 $M_s=10^{-3}\,\mathrm{GeV}$, $\lambda=1$, $\mu=1\,\mathrm{keV}$
 (depth scale $V_0=15.9\,\mathrm{meV}$, surface depth
 $|V_s(0)|=15.1\,\mathrm{meV}$), with three bound states within \aa ngstroms of
 the surface.   Wavefunctions in panels~(a,b,d) have arbitrary vertical scaling.
  In panels~(a,b), each is offset by its eigenenergy.
 (b)~The gravitational zone of the same point, the Airy quartet
 is unchanged at the peV scale, and the three layer states far
 below it take the lowest indices in the sorted ladder.
 (c)~Reduced-layer energies $|E^{\rm layer}_n|$ versus $\mu$ for
 the model of panel~(a). Labels above the panel give
 representative layer widths.  Crosses are full
 layer-plus-gravity solves at each decade (at $1\,\mathrm{keV}$
 at relaxed spacing), tracking the reduced branches across
 roughly eight orders of magnitude in energy. The near-threshold
 fourth branch (grey) appears as a negative level of the full
 problem only where its binding exceeds the gravitational tilt
 (see text).  (d)~At the disputed-strip point
 $M_s=10^{-5}\,\mathrm{GeV}$, $\lambda=10^{2}$, the layer binds
 nothing, and the zero-energy solution approaches
 $\psi\propto z-a_{\rm s}$ with scattering length $a_{\rm s}=-0.19\,\mathrm{nm}$
 (dashed).}
 \label{fig:qbounce-states}
\end{figure*}

Figure~\ref{fig:qbounce-states} illustrates the spectral distinction
  at $\mu=1\,\mathrm{keV}$.  At
  $M_s=10^{-3}\,\mathrm{GeV}$, $\lambda=1$, three surface-bound
  states at meV energies lie below a nearly unchanged gravitational
  ladder [panels~(a,b)].  Under the adopted numbering, these become
  the lowest three levels, giving $E_{13}=7.62\,\mathrm{meV}$,
  far outside the measured peV-scale window.  At
  $M_s=10^{-5}\,\mathrm{GeV}$, $\lambda=10^2$, the layer instead
  supports no bound states [panel~(d)].  Its scattering length,
  $a_{\rm s}=-0.190\,\mathrm{nm}$, produces only a common shift of the
  gravitational levels at leading order, leaving their transition
  energies unchanged.  The contrasting results therefore arise
  from the appearance of surface-bound states and the resulting
  change in level numbering, while the gravitational ladder itself
  is almost unchanged.

  Figure~\ref{fig:qbounce-states}(c) varies $\mu$ at fixed
  $M_s$ and $\lambda$. The dimensionless strength
  \begin{equation}
   S=\frac{\Q m_n^2}{2\lambda M_s^2}
   \label{eq:layer-strength-identity}
  \end{equation}
  depends on $\mu$ only through $\Q$, which varies little
  along this scan. If $w_s$ were also fixed, the dimensionless
  eigenvalues $\varepsilon_n$ would therefore remain nearly
  constant. The physical energies
  $E_n^{\rm layer}=E_u\varepsilon_n$ would then scale as
  $\mu^2$, since $E_u\simeq\mu^2/m_n$ away from symmetry
  restoration. This explains the main trend of the three
  deeply bound levels. Changes in $w_s$ produce departures
  from this scaling and remove the shallow fourth inner
  bound state by $1\,\mathrm{keV}$. For this fourth state,
  weak binding also makes the neglect of gravity less reliable.

Binding requires the attractive well to overcome the kinetic-energy
cost of localizing a neutron next to the hard wall. The parameter
$S$ compares the well depth with this kinetic-energy scale.
When the field vanishes at the mirror ($w_s=1$), the first four
binding thresholds are $S=1/2,3,7.5,14$. A nonzero surface field
changes these thresholds. At the excluded $1\,\mathrm{keV}$
worked point, $w_s\simeq0.646$ and only three states remain bound,
whereas the weaker well at the allowed point supports none.

  For the $1\,\mathrm{keV}$ constraint grid, binding of the isolated layer is a useful diagnostic but is not the exclusion criterion. Near the
  first threshold we retain energy and gravity in the inner equation
  and match to the decaying outer Airy solution, and each cell is
  decided by spectral bounds on the transition energies under the
  adopted level numbering. A nonbinding layer can still be excluded. At $w_s=1$ and $S=0.49999<1/2$ the scattering length is
  $-20.93\,\mu\mathrm{m}$ and the gaps are $2.17669$ and
  $2.97738\,\mathrm{peV}$, both outside the measured windows. The strip so
  added is narrower than the grid spacing, so no displayed exclusion bound changes.

  Table~\ref{tab:qbounce-levels} collects the worked-point spectra
  across the three regimes. At $\mu=1\,\mathrm{meV}$, the broad
  well mainly lowers the gravitational ladder, while its spatial
  variation changes the transition energies, so negative energies
  alone do not imply surface localization. At $0.1$ and
  $1\,\mathrm{eV}$, increasingly localized surface states coexist
  with a modified gravitational ladder. 

  \begin{table*}[tbp]
  \caption{Five lowest levels and transition energies at the exact
  quoted parameter points. The first four rows lie inside our
  sphere exclusion regions in
  Fig.~\ref{fig:constraints-symmetron}; the final row lies inside
  the published exclusion region but outside ours.
  ``Full'' denotes diagonalization on the production mesh;
  ``reduced'' denotes a reconstructed two-zone spectrum.
  Energies use the convention $V_s(\infty)=0$ and are listed
  in increasing order.
  The transitions are $E_{13}=E_3-E_1$ and $E_{14}=E_4-E_1$,
  with measured windows $(1.9114,1.9330)$ and
  $(2.6726,2.7022)\,\mathrm{peV}$, respectively.}
 \label{tab:qbounce-levels}
 \begin{ruledtabular}
 \begin{tabular}{lllllllll}
 $\mu$ & $M_s$ [GeV] & $\lambda$ & $V_0$ & $w_s$ & $S$ & method & five lowest levels [peV] & $E_{13}$, $E_{14}$ [peV] \\
 \hline
 $1\,\mathrm{meV}$ & $10^{-1}$ & $10^{-4}$ & $95.4\,\mathrm{peV}$ & $1.00$ & --- & full & $-93.9,\,-92.6,\,-91.5,\,-90.5,\,-89.5$ & $2.35$, $3.36$ \\
 $0.1\,\mathrm{eV}$ & $10^{-3}$ & $1$ & $143\,\mathrm{peV}$ & $1.00$ & $14.9$ & full & $-93.1,\,-42.8,\,-11.6,\,0.915,\,2.52$ & $81.5$, $94.0$ \\
 $1\,\mathrm{eV}$ & $10^{-3}$ & $1$ & $15.9\,\mathrm{neV}$ & $1.00$ & $14.9$ & full & $-10344,\,-4777,\,-1336,\,-15.7,\,1.59$ & $9.01\times10^{3}$, $1.03\times10^{4}$ \\
 $1\,\mathrm{keV}$ & $10^{-3}$ & $1$ & $15.9\,\mathrm{meV}$ & $0.646$ & $14.9$ & reduced & $(-7.99,\,-2.84,\,-0.370)\times10^{9},\,1.41,\,2.46$ & $(7.62,\,7.99)\times10^{9}$ \\
 $1\,\mathrm{keV}$ & $10^{-5}$ & $10^{2}$ & $0.159\,\mathrm{meV}$ & $0.996$ & $0.15$ & reduced & $1.41,\,2.46,\,3.32,\,4.08,\,4.78$ & $1.915$, $2.676$ \\

 \end{tabular}
 \end{ruledtabular}
\end{table*}

This appendix presents the methods used to calculate the ordered
qBOUNCE spectra within the hard-wall approximation and tests their
numerical convergence at representative worked points. Finite mirror
penetration is not included. Nor do these calculations determine whether
the surface-bound states actually participate in the measured
transitions. That requires a separate analysis of state preparation,
transition strengths, and detection, which we leave to future study.

\clearpage   
\bibliography{references}

\end{document}